\documentclass[12pt,preprint]{aastex}

\newcommand{\mdot}{M$_{\odot}$ yr$^{-1}$}
\newcommand{\ldot}{L$_{\odot}$}
\newcommand{ \um}{$\mu$m~}
\newcommand{ \ums}{$\mu$m}
\def\kmsMpc{\ifmmode {\rm\,km\,s^{-1}\,Mpc^{-1}}\else
    ${\rm\,km\,s^{-1}\,Mpc^{-1}}$\fi}

\shorttitle{ Far Infrared Luminosities for Dusty Quasars}
\shortauthors{Weedman and Sargsyan}

\begin{document}

\title{Dusty Quasars at High Redshifts}

\author{ Daniel Weedman\altaffilmark{1} and Lusine Sargsyan\altaffilmark{1}}

\altaffiltext{1}{Cornell Center for Astrophysics and Planetary Science, Cornell University, Ithaca,
NY 14853, USA; dweedman@astro.cornell.edu}

\begin{abstract}

A population of quasars at z $\sim$ 2 is determined based on dust luminosities $\nu L_{\nu}$(7.8 \ums) that includes unobscured, partially obscured, and obscured quasars.  Quasars are classified by the ratio $\nu L_{\nu}$(0.25 \ums)/$\nu L_{\nu}$(7.8 \ums) = UV/IR, assumed to measure obscuration of UV luminosity by the dust which produces IR luminosity.  Quasar counts at rest frame 7.8 \um are determined for quasars in the Bo{\"o}tes field of the NOAO Deep Wide Field Survey using 24 \um sources with optical redshifts from the AGN and Galaxy Evolution Survey (AGES) or infrared redshifts from the $Spitzer$ Infrared Spectrograph. Spectral energy distributions are extended to far infrared wavelengths using observations from the $Herschel$ Space Observatory Spectral and Photometric Imaging Receiver (SPIRE), and new SPIRE photometry is presented for 77 high redshift quasars from the Sloan Digital Sky Survey.  It is found that unobscured and obscured quasars have similar space densities at rest frame 7.8 \ums, but the ratio $L_{\nu}$(100 \ums)/$L_{\nu}$(7.8 \ums) is about three times higher for obscured quasars compared to unobscured, so that far infrared or submm discoveries are dominated by obscured quasars.  Quasar source counts for these samples are determined for comparison to the number of submm sources that have been discovered with the SCUBA-2 camera at z $\sim$ 2 using the $L_{\nu}$(100 \ums)/$L_{\nu}$(7.8 \ums) results together with the Bo{\"o}tes 7.8 \um counts, and we find that only $\sim$ 5\% of high redshift submm sources are quasars, including even the most obscured quasars. Illustrative source counts are predicted to z = 10, and we show that existing SCUBA-2 850 \um surveys or 2 mm surveys with the Goddard-IRAM Superconducting 2 Millimeter Observer (GISMO) survey camera should already have detected sources at z $\sim$ 10 if quasar and starburst luminosity functions remain the same from z = 2 until z = 10. 

\end{abstract}

\keywords{quasars: general---
        infrared: galaxies ---
  	galaxies: active---
	galaxies: high redshift---
 	galaxies: evolution---
	galaxies: starburst
	}

\section{Introduction}

As infrared and submillimeter observational capabilities developed over the past two decades, the census of dusty sources in the extragalactic universe increased dramatically for redshifts z $\ga$ 2.  Such sources are crucial for gaining a full description of formation and evolution for galaxies and quasars in the universe, because optically derived surveys are subject to severe selection effects when the rest frame ultraviolet is affected by dust extinction. High redshift, dusty sources unknown from optical surveys were initially found at z $\ga$ 2 in 850 \um surveys with the Submillimeter Common User Bolometric Array (SCUBA) camera \citep{sma97,cha05}, then with 24 \um surveys and follow-up spectroscopy with the $Spitzer$ Space Telescope \citep{hou05,yan05}, more recently \citep{eis12} among 12 \um and 22 \um sources found by the Wide-Field Infrared Survey Explorer (WISE) and in far infrared surveys \citep{cas12,dow14} with the $Herschel$ Space Observatory Spectral and Photometric Imaging Receiver (SPIRE).

Based initially on the $Spitzer$ surveys, a population of "Dust Obscured Galaxies" (DOGs) was defined \citep{dey08}, and a scenario was developed to explain their formation and evolution.  To summarize simply, the assembly of the earliest massive galaxies is characterised by extensive dust formation arising in the short lived, initial stellar populations.  The remnants of these populations lead to formation of supermassive black holes which power luminous active galactic nuclei (AGN) observed as dust obscured quasars.  Eventually, radiation pressure from the quasars expels the dust, leading to the optically observable quasars whose apparent luminosity peaks at z $\sim$ 2 \citep{hop08,nar10}. 

This scenario means that observational determinations of the formation and evolution of the earliest massive galaxies and supermassive black holes must use primarily the observed reradiation from the obscuring dust, which can be achieved only with infrared through millimeter wavelengths.  Particularly crucial is the determination of which sources have their dust luminosity arising because of quasars, and which sources arise from star formation within luminous starbursts. 

In the present paper, we summarize the dusty quasars in observed quasar populations already known at redshifts z $\sim$ 2 where their discovery has been most complete, and then project the number that should be seen at z $\sim$ 10 using submillimeter or millimeter surveys.  Our motive is to provide comparisons to the rapidly improving sensitivity of submillimeter and millimeter surveys with SCUBA-2 at 450 \um and 850 \um \citep{gea13,ros13,bar14} and with the Goddard-IRAM Superconducting 2 Millimeter Observer (GISMO) survey camera \citep{sta14},  together with the capability of measuring high redshifts of dusty sources with submillimeter interferometers.  Already, for example, the [CII] 158 \um emission line has been measured in sources with 4 $<$ z $<$ 7.1 \citep{huy13,wan13,car13,ban15}.    

We determine empirical luminosity functions and quasar counts at z $\sim$ 2 for all categories of dusty quasars, including fully obscured, partially obscured, and unobscured quasars, defining the amount of obscuration by the ratio $\nu L_{\nu}$(0.25 \ums)/$\nu L_{\nu}$(7.8 \ums) = UV/IR.  Although this is an observational classification independent of the interpretion, we describe the ratio as a measure of ultraviolet obscuration by dust and use it to define the three categories of quasars. For all categories, luminosity functions and source counts are normalized to the dust continuum luminosity $\nu L_{\nu}$(7.8 \ums) at rest frame 7.8 \um to minimize effects of extinction for optically obscured quasars. This particular wavelength is used because it is a localized spectral maximum for quasars heavily absorbed by the 9.7 \um silicate feature and allows a uniform comparison between obscured quasars with large extinction and silicate absorption, and the unobscured optical quasar samples with little extinction and silicate emission.  Among AGN, this measure of dust luminosity correlates well with hard X-ray luminosity, black hole mass, and high ionization emission line luminosity \citep{wee12}. The obscured quasars with 9.7 \um silicate absorption, unknown from optical surveys, are those discovered in surveys with the $Spitzer$ Infrared Spectrograph (IRS, Houck et al. 2004).   

To predict detections at longer wavelengths, we determine the far infrared spectral energy distributions (SEDs) of obscured and unobscured quasars using observations with SPIRE.  This includes previously published results for obscured quasars together with our own new SPIRE photometry of 77 unobscured quasars from the quasar catalog \citep{sch10} of the Sloan Digital Digital Sky Survey (SDSS, Gunn et al. 1998).  

These empirical results for dusty quasars are used to produce quasar source counts for comparison to the number of sources that have been discovered with SCUBA-2 at z $\sim$ 2.  To illustrate an example of future discovery possibilities, we determine the number of quasars that should be seen for 9.5 $<$ z $<$ 10.5 with SCUBA-2 and GISMO if quasar luminosity functions stay constant for z $>$ 2.  Eventual comparison of this prediction with observations will allow a measure of whether the formation rate of luminous quasars and the mix of quasars and starbursts changed between 2 $\la$ z $\la$ 10.  We determine luminosities throughout using H$_0$ = 74 \kmsMpc \citep{rie11}, $\Omega_{M}$=0.27, and $\Omega_{\Lambda}$=0.73.

\section{Dusty Quasar Populations}

Quasar surveys at optical wavelengths naturally favor those quasars which are luminous in the rest frame ultraviolet and which have broad emission lines for classification and redshift measurement.  These "type 1" quasars dominate classical samples \citep{car78,lew79,osm82,sch83,mar84,boy88} and extensive recent surveys such as the SDSS quasar catalog \citep{sch10} and the AGN and Galaxy Evolution Survey (AGES, Kochanek et al. 2012). Type 1 quasars are presumed to show unobscured quasars whose intrinsic ultraviolet and emission line luminosities are not affected by dust extinction. By contrast, extensive observational studies of type 2 quasars \citep{wil00,ale03,zak04,mar06,hic07} are interpreted as showing partially obscured quasars, in which the broad line region and intrinsic ultraviolet continuum are not observed.  These interpretations arise as an extension of the Seyfert 1 and Seyfert 2 active galactic nucleus (AGN) classifications, originally defined spectroscopically based on the presence or absence of broad hydrogen emission lines \citep{kw74} and subsequently interpreted within the "unified theory" as arising from orientation effects that could obscure the broad line region \citep{ant93}. 

Our goal is the definition of a quasar population that is not biased by extinction effects, ranging from unobscured quasars through the DOGs population.  To achieve this, we classify quasars quantitatively based on the ultraviolet to infrared luminosity ratio.  Following \citet{var14}, we use the rest frame ratio UV/IR = $\nu L_{\nu}$(0.25 \ums)/$\nu L_{\nu}$(7.8 \ums). This parameter is chosen because of reasons given earlier for $\nu L_{\nu}$(7.8 \ums), and because $\nu L_{\nu}$(0.25 \ums) is determined spectroscopically for SDSS quasars \citep{she11}.  Categories are chosen that cover UV/IR for all quasars, and we assume based on previous work that this ratio is controlled primarily by the amount of extinction that suppresses $\nu L_{\nu}$(0.25 \ums).  In the following discussions, we group quasars into three categories based on empirical determinations of UV/IR: obscured quasars with log UV/IR $<$ -1.8, partially obscured with -1.8 $<$ log UV/IR $<$ 0.2, and unobscured quasars with log UV/IR $>$ 0.2. For luminosity functions and quasar counts, we compare these categories within a specific redshift interval near z $\sim$ 2 for which surveys for all categories are most complete because infrared-derived redshifts have been determined for obscured quasars, independent of optical detections and dust extinction. 

An infrared spectroscopic classification based on the 9.7 \um silicate feature also correlates well with obscured and unobscured classifications and the UV/IR ratio. The presence of silicate absorption means there must be cooler dust between the observer and the hotter dust responsible for the infrared continuum; sources with the smallest values of UV/IR, the DOGS, have measurable redshifts only because of strong silicate absorption.  Observing silicate emission means that the hotter side of the clouds is directly observed, implying little extinction.  This interpretation is consistent with observations of silicate strengths and the correlation with type 1 and type 2 AGN classifications \citep[e.g.][]{hao05,ima07,hao07,wee12} and with dusty torus models \citep{shi06,ra11,efs14}. 

DOGs have also been extensively studied in $Spitzer$ photometric surveys at various wavelengths and large samples of obscured and unobscured quasars were defined using colors from the $Spitzer$ Infrared Array Camera (IRAC; Fazio et al. 2004). A comprehensive summary of the history and definition of these photometric samples is in \citet{che15}.  Sources with power law continua extending through the IRAC bands are interpreted as AGN \citep{brn06,don07,bus09,mel12} and sometimes called "power law DOGS".  These are contrasted to sources having a photometric peak within the IRAC bands, sometimes called "bump" DOGS, which is interpreted as arising from the rest frame 1.8 \um absorption in stellar atmospheres \citep{si99}.  The AGN DOGs overlap in characteristics with many of the Compton thick, obscured X-ray sources \citep{brn08,pol08,fio08,bau10}.  As verified below, the AGN DOGS generally show the 9.7 \um silicate absorption feature when IRS spectra are available.  Conversely, sources chosen photometrically from $Spitzer$ surveys as "bump" sources consistently show PAH features in IRS spectra \citep{wee06c,far08,des09,fio10}.

All individual quasars or AGN which are discussed in this paper are summarized in Figure 1 showing dust luminosities $\nu L_{\nu}$(7.8 \ums) to illustrate the range of redshifts and mid-infrared dust luminosities encompassed in our analysis.  Sources in this Figure are classified based on the spectroscopic silicate criterion, so unobscured quasars are those with silicate emission and obscured quasars are those with silicate absorption.  The systematic differences in $\nu L_{\nu}$(7.8 $\mu$m) between the two samples of high redshift quasars (SDSS/WISE unobscured, $Spitzer$ IRS obscured) arise primarily from differences in survey areas.  The SDSS/WISE sample covers a large sky area of $>$ 10,000 deg$^{2}$, whereas the obscured quasars arise only within $\sim$ 10 deg$^{2}$, so the smaller area survey does not reach the rare but more luminous sources within the larger survey.  This figure also shows that the highest infrared luminosities continue to the highest redshifts observed, with no turndown at any redshift yet found, a result described in more detail for SDSS/WISE quasars in \citet{var14}.

\begin{figure}
\figurenum{1}
\includegraphics[scale= 1.0]{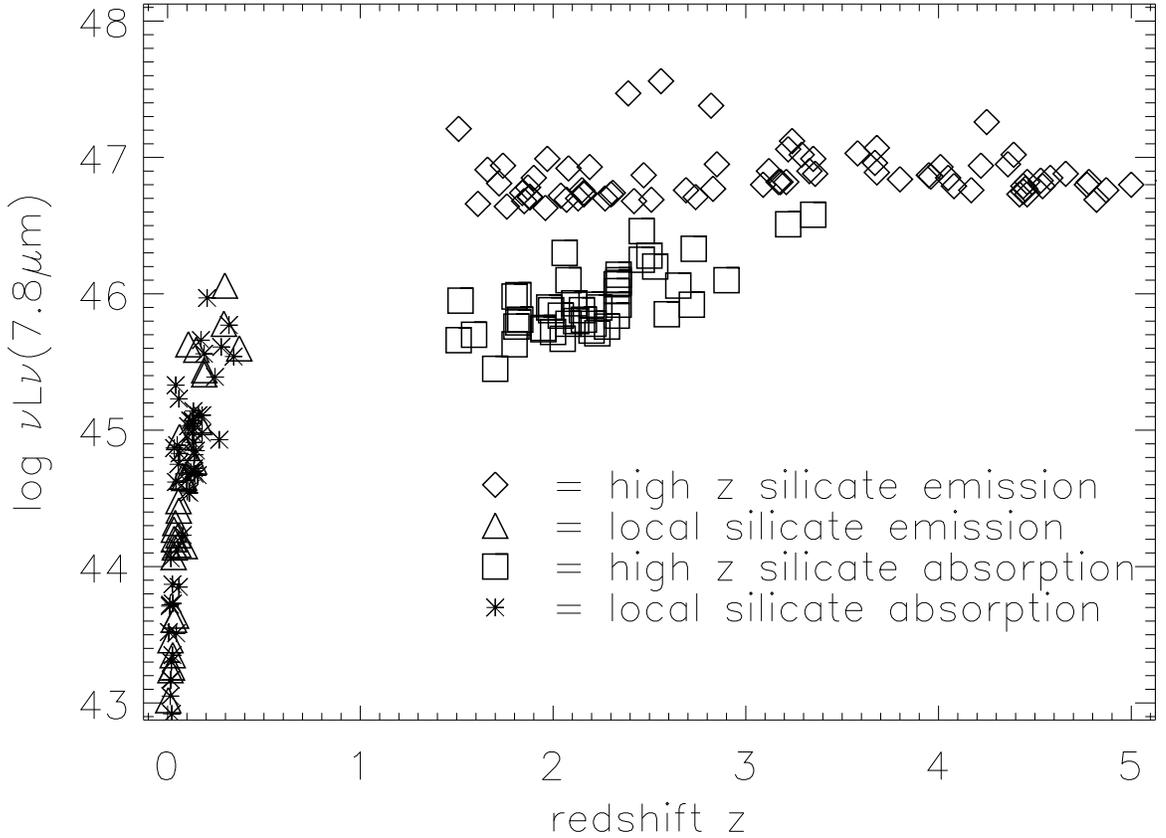}
\caption{The $\nu L_{\nu}$(7.8 \ums) (erg s$^{-1}$) distribution with redshift of individual sources used in this paper for determining ratios of far infrared to $\nu L_{\nu}$(7.8 \ums) luminosity.  Triangles are low redshift silicate emission AGN with far infrared luminosities from IRAS, asterisks are low redshift silicate absorption AGN with far infrared luminosities from IRAS, diamonds are high redshift SDSS/WISE quasars with new SPIRE photometry in Table 2, and squares are high redshift silicate absorption quasars discovered by $Spitzer$ IRS having published SPIRE photometry \citep{mel12,saj12}.  For SDSS/WISE detections, an empirical IRS template is used to transform observed frame $f_{\nu}$(22 \ums) to rest frame $\nu L_{\nu}$(7.8 \ums). As explained in the text, the total infrared dust luminosity $L_{IR}$ is empirically determined from $\nu L_{\nu}$(7.8 \ums) as log [$L_{IR}$/$\nu L_{\nu}$(7.8 \ums)] = 0.51 in low redshift AGN with silicate emission, log [$L_{IR}$/$\nu L_{\nu}$(7.8 \ums)] = 0.80 in low redshift AGN with silicate absorption, log [$L_{IR}$/$\nu L_{\nu}$(7.8 \ums)] = 0.41 for the high redshift silicate emission quasars, and log [$L_{IR}$/$\nu L_{\nu}$(7.8 \ums)] = 0.68 for the high redshift silicate absorption quasars.  }

\end{figure}

\subsection{Obscured Quasars}

The original definition of DOGs in \citet{dey08} assumed that small UV/IR ratios arise because of dust extinction, a conclusion based primarily on the presence of silicate absorption in the original DOG quasar samples as proof of intervening dust.  Subsequent study confirmed that extinction was indeed the best explanation for the general DOG population, rather than intrinsic differences in SEDs \citep{pen12}. These obscured DOG quasars were first found using $Spitzer$ IRS spectroscopy \citep{hou05,yan05} that discovered optically faint quasars having redshifts measureable only from the 9.7 \um silicate absorption feature.  Subsequently, the DOGs were defined by Dey et al. as having observed infrared to optical flux density ratios $f_{\nu}$(24 \ums)/$f_{\nu}$($R$) $>$ 1000, or $R$ - [24] $>$ 14 (Vega magnitudes).  Adopting that a [24] magnitude of zero corresponds to 7.3 Jy, the DOG definition means that any source having $f_{\nu}$(24 \ums) $>$ 1 mJy and $R$ $>$ 23.7 would be a DOG.  In the quantitative counts of DOGS discussed below, we determine completeness corrections for obscured quasars in the DOG surveys based on $R$ $>$ 24.  We define these magnitudes as "optically faint".  

\begin{figure}
\figurenum{2}
\includegraphics[scale= 1.0]{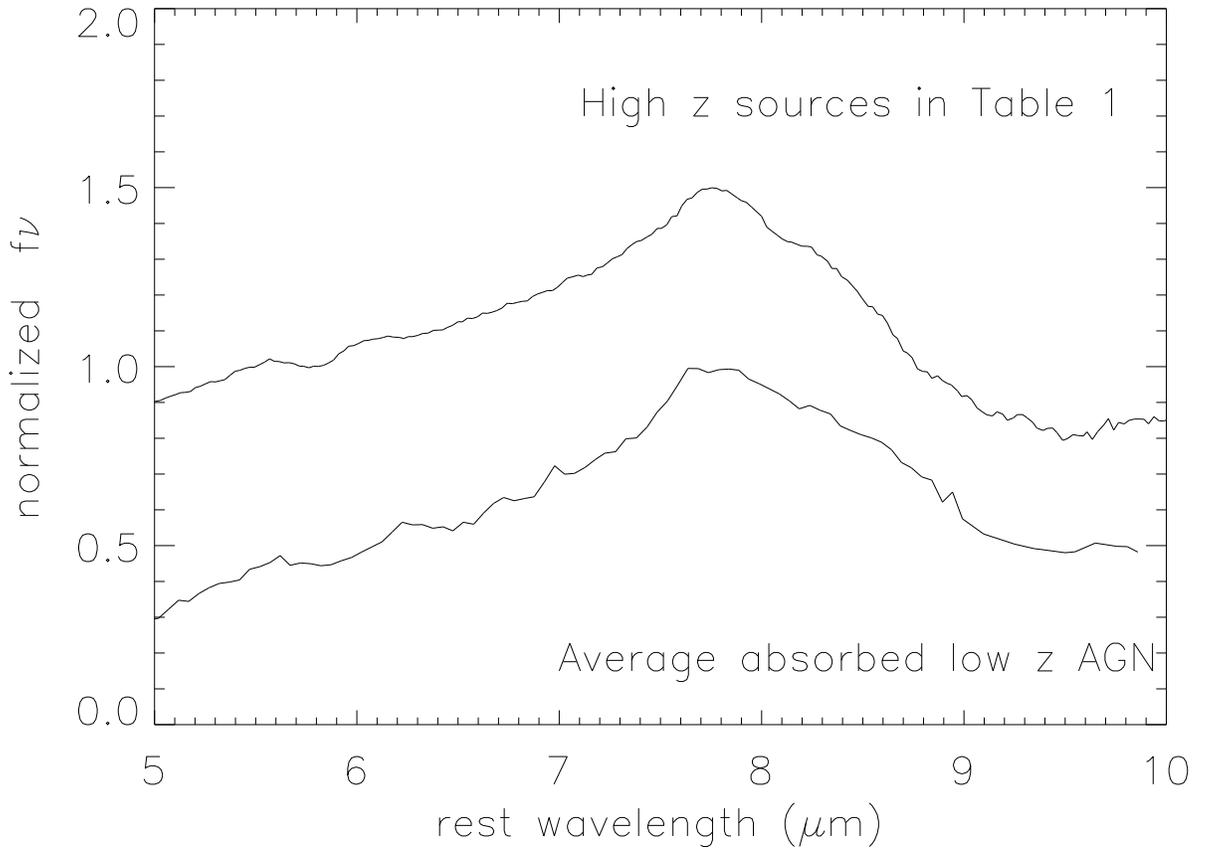}
\caption{Upper spectrum is normalized average rest frame spectrum of all silicate absorption quasars used in this paper (sources with z $>$ 1.5 discovered with $Spitzer$ IRS given in Table 1). Bottom spectrum is average observed rest frame spectrum of 65 silicate absorbed low redshift AGN from \citet{sar11} used as comparisons for far infrared SEDs. Spectra are normalized to peak $f_{\nu}$(7.8 $\mu$m) and displaced by 0.5 units of $f_{\nu}$. } 

\end{figure}

The obscured quasars defining the DOGs were initially found among sources identified in 24 \um surveys using the $Spitzer$ MIPS instrument \citep{rie04}, primarily of the Bo{\"o}tes field of the NOAO Deep Wide Field Survey (NDWFS, Jannuzi and Dey 1999) and the $Spitzer$ First Look Survey (FLS, Fadda et al. 2006).  To assemble our present summary of obscured quasars, we have utilized all sources with IRS spectra within the Bo{\"o}tes field \citep{hou05,wee06,mel12} and the FLS field \citep{yan07,saj07,das09,wee06b} meeting the photometric criteria of $f_{\nu}$(24 \ums) $>$ 1 mJy and $R$ $>$ 24.  Although redshifts and spectroscopic identifications of the silicate absorption feature had previously been identified in most cases, we reexamined all spectra using the improved spectral extractions in the CASSIS spectral atlas \citep{leb11}\footnote{http://cassis.sirtf.com. The Cornell Atlas of Spitzer IRS Spectra (CASSIS) is a product of the Infrared Science Center at Cornell University.}.  We also measured rest frame $f_{\nu}$(7.8 \ums) in all sources from the CASSIS spectra. 

Our total sample of obscured quasars is given in Table 1, with the Bo{\"o}tes sources reproduced from \citet{var14}. Because the $Spitzer$ quasars derive initially from 24 \um surveys, there is a strong redshift selection for silicate absorption sources when the 7.8 \um continuum peak is near 24 \um in the observed frame.  To accommodate this selection, we describe luminosity functions and source counts only within the range 1.8 $<$ z $<$ 2.4, although all obscured quasars with z $>$ 1.5 meeting our photometry definitions are listed in Table 1.  Published photometry from $Herschel$ SPIRE is also included in this Table for the discussion of far infrared SEDs which follows below.

Other than the DOG criterion, the most important selection to be applied is to assure that we identify DOGs which are powered by the AGN of a quasar, without having a significant contribution to dust luminosity from a starburst component.  We base this decision also on an IRS spectroscopic criterion.  Many studies have shown that the strength of the polyclyclic aromatic hydrocarbon (PAH) features with rest frame wavelengths 6 \um $<$ $\lambda$ $<$ 12 \um are a measure of the starburst component \citep[e.g.][]{gen98,lau00,bra06,des07,vei09,tom10,wu10,sar11,sti13}.  For high redshift sources, the PAH feature used is at 6.2 \ums.  A classification generally adopted is that any source with rest frame EW(6.2 \ums) $<$ 0.1 \um is dominated by AGN luminosity. 

Because of the poor S/N of many sources in our present study, it is not realistic to apply a rigorous criterion for EW(6.2 \ums) to each source to determine the classification, but no source included in Table 1 has a measurable 6.2 \um feature that exceeds the spectral noise, and the upper limits are significantly smaller than 0.1 \ums.  This is illustrated by the average spectrum of all sources in Table 1, shown in Figure 2, for which the EW(6.2 \ums) = 0.018 \ums.  This small EW is evidence that the sample is indeed dominated by ``pure" AGN.   The average spectrum also illustrates the 7.8 \um peak flux density that is measured and shows for comparison the low redshift, silicate absorption AGN used as local analogues. The distinctive difference between an obscured quasar with silicate absorption and a source with PAH emission is illustrated below in section 5.1.  We note also that 26 of the Bo{\"o}tes sources in Table 1 which define our sample of obscured quasars are photometrically classified by \citet{mel12}, and 23 of 26 are "power law" DOGS, with only 3 classed as "bump" sources.

The limiting UV/IR for obscured quasars cannot be determined using monochromatic wavelengths because rest frame ultraviolet flux densities are measured only with broad band $R$ and $I$ filters, and which filter is closer to rest frame 0.25 \um depends on redshift; effective wavelengths are $\sim$ 0.65 \um and 0.80 \um.  For the redshift interval 1.8 $<$ z $<$ 2.4 we use, the observed frame wavelength for rest frame 0.25 \um is 0.7 \um $<$ $\lambda$ $<$ 0.85 \ums.  The magnitudes of the brightest IRS obscured quasars are $\sim$ 24 in either filter, corresponding to 0.77 or 0.61 $\mu$Jy for $R$ or $I$.  Taking the average as representing the brightest obscured quasar ($R$ = 24) and comparing to the faintest $f_{\nu}$(7.8 \ums) in Table 1 ($\sim$ 1.5 mJy) yields a limiting log [$\nu L_{\nu}$(0.25 \ums)/$\nu L_{\nu}$(7.8 \ums)] $<$ -1.8.  All obscured quasars have values of UV/IR smaller than this.

\subsection{Unobscured Quasars}

The largest sample of quasars that are unobscured, optically bright, and having measurable $\nu L_{\nu}$(7.8 \ums) are those within the SDSS \citep{sch10}. Their dust luminosities $\nu L_{\nu}$(7.8 \ums) can be determined using the WISE 22 \um photometry for SDSS quasars together with a template spectrum to transform observed frame 22 \um to rest frame 7.8 \um \citep{wee12,var14}.  The template is determined from IRS spectra of SDSS quasars \citep{deo11} which are characterized by silicate emission, indicating that these quasars are unobscured. The $\nu L_{\nu}$(0.25 \ums) of SDSS quasars are tabulated \citep{she11}, so UV/IR can be determined. The results in Vardanyan et al. show that all SDSS quasars in the redshift interval we use have log UV/IR $>$ 0.2, which defines our classification of unobscured quasars. 

For our analysis, the most useful sample of unobscured quasars is the fainter sample of type 1 optical quasars available in the AGES survey \citep{koc12}, which have similar spectra and UV/IR ratios to the SDSS quasars.  This covers the same Bo{\"o}tes survey field as the obscured IRS quasars but reaches fainter dust luminosities $\nu L_{\nu}$(7.8 \ums) than SDSS because AGES utilizes the Bo{\"o}tes photometry going to 0.3 mJy at 24 \um whereas the SDSS/WISE quasars only reach 2 mJy at 22 \um.

\subsection{Partially Obscured Quasars}

The partially obscured quasars are, by definition, intermediate between the obscured DOGs and the unobscured type 1 quasars defined above. This defines their -1.8 $<$ log UV/IR $<$ 0.2.  Observationally, this implies that partially obscured quasars are found primarily within samples classified as type 2 quasars. A good example of this is among the type 2 quasars within the AGES samples.  For example, from Figure 7 in \citet{hic07}, the sample of type 1 has median $R$ = 21, but type 2 has median $R$ = 23, and the faint limits of the two samples are also shifted by $\sim$ 2 mag.  For similar distributions of $f_{\nu}$(24 \ums) among type 1 and type 2, this result demonstrates that median log UV/IR for the type 2 samples is systematically smaller by about 0.8 than for type 1, so that the type 2 represent a partially obscured sample with representative log UV/IR $\sim$ -0.6.

\section{Far Infrared Luminosities of Dusty Quasars}

A major purpose of this paper is to determine the source counts expected at submillimeter and millimeter wavelengths for the full quasar population including all classes of obscuration.  These results are enabled by observations of quasars with the SPIRE instrument \citep{gri10} on the $Herschel$ Space Observatory \citep{pil10} to determine far infrared luminosities.  In this section, we compare results for unobscured quasars using new observations to results for obscured quasars using previously published results in the Bo{\"o}tes and FLS fields. 

Although we are determining the far infrared luminosities of sources classified spectroscopically as quasars, we note the extensive previous studies summarized in \citet{che15} that attribute the far infrared luminosity from many AGN and quasars to dust reradiation from a starburst component.  These conclusions rely on SED template libraries which show that starbursts have stronger far infrared luminosity than AGN \citep{cha01, dal02, asf10, elb11, war11}.  The classification of infrared SEDs in these libraries arises primarily from the PAH features or from emission line ratios that correlate with PAH features. For these reasons, using the absence of PAH features to define obscured AGN or quasars is consistent with the approach of previous studies based on SEDs.  For sources without PAH features, there is no spectroscopic evidence to attribute the far infrared luminosity to a starburst.  More detailed efforts to deconvolve starburst and AGN far infrared components based on SEDs also show good correlations between PAH strength and the starburst luminosity component \citep{fel13, hil14}. In either case, it could never be proven based on any criterion whether a source is starburst or AGN if sources are so obscured that all spectroscopic indicators from emission line ratios or PAH strengths are hidden \citep{pol08,far07,des07,kir12,lei14}.

\subsection{Observations with SPIRE of SDSS/WISE Unobscured Quasars}

To assemble far infrared luminosities of luminous, unobscured quasars at high redshifts, we selected sources from the SDSS quasar catalog for new observations with SPIRE photometry. Our selection of unobscured sources proposed for SPIRE cycle 2 observations was made by comparing the SDSS quasar catalog \citep{sch10} with the first WISE data release \citep{wri10} using a criterion $\la$ 3~$\arcsec$ for source identification.  This resulted in 9424 SDSS/WISE quasars detected at 22 \um.  The observed fluxes were scaled to $f_{\nu}$(rest frame 7.8 \micron) by using SDSS redshifts combined with an empirical spectral template we determined using IRS spectra of type 1 AGN and SDSS quasars; this template is illustrated and defined in \citet{wee12} and \citet{var14}. Initially, we chose the most infrared luminous 25 quasars in each redshift interval of 0.5 for 1.5 $<$ z $<$ 5 which resulted in 175 sources. Of these 175, 77 were successfully observed with SPIRE in program dweedman-OT2 before the $Herschel$ mission ended.   

Our observations of individual sources were made with the SPIRE small map mode\footnote{http://herschel.esac.esa.int/Docs/SPIRE/html/spire-handbook.html}.  Photometry was analyzed using the $Herschel$ Interactive Processing Environment (HIPE) v11.1.0 and the SPIRE Small Map Mode User Reprocessing Script\footnote{http://herschel.esac.esa.int/hipe/}. Photometry of sources was done with SUSSExtractor point source extraction \citep{sav07}.  These techniques were used so that our results were derived in similar fashion to those of HerMES, which we use below for comparison to obscured quasars already published.  The signal to noise (S/N) threshold was set at 3, and full width half maximum (FWHM) for the point spread function (PSF) were taken as 18.2\arcsec, 24.9\arcsec~ and 36.3\arcsec~ for 250 \ums, 350 \ums, and 500 \ums.  A detection is assumed to be real if the distance between the SPIRE source and the SDSS coordinate is $<$ 18\arcsec~for 250 \ums, $<$ 25\arcsec~for 350 \ums, and $<$ 37\arcsec~for 500 \um (i.e. distances of one FWHM of the PSF).  If no source is detected within these criteria, an upper limit of 25 mJy is assumed at all wavelengths.  Results of the new photometry are in Table 2.  The photometry we report gives the measured fluxes of sources at the positions listed.  We apply no statistical corrections for faint, underlying background sources that might artifically boost the observed fluxes \citep[e.g.][]{bet12}, because the flux limits we use exceed by 4 $\sigma$ the background confusion noise \citep{ngu10}.

\subsection{Comparisons of SEDs for Obscured and Unobscured Quasars}

Far Infrared luminosities of obscured quasars from the Bo{\"o}tes and FLS survey fields are given in Table 1.  These were determined by \citet{mel12} for Bo{\"o}tes and \citet{saj12} for FLS using the $Herschel$ Multi-tiered Extragalactic Survey with SPIRE (HerMES; Oliver et al. 2010). 

Our SEDs are all normalized to $L_{\nu}$(7.8 $\mu$m), because this wavelength defines the luminosity functions and quasar counts we use. Figures 3 and 4 show the SEDs of the unobscured and obscured quasar samples, as determined by the SPIRE observations.  For comparison with the luminous, high redshift quasars, we also include low redshift AGN which have both IRS spectra for classification and measures of $f_{\nu}$(7.8 \ums) as well as photometry with the Infrared Astronomical Satellite (IRAS) to give fluxes at $\sim$ 100 \um.  These AGN are listed in \citet{sar11}.  

One of our goals for the SPIRE observations is to test the simple expectations of the unified model, explaining obscured and unobscured quasars as differing only in orientation of a dusty torus that can obscure ultraviolet luminosity.  In this interpretation, the infrared SEDs should not show differences between obscured and unobscured quasars if extinction does not affect the infrared.  Detailed considerations of radiative transfer effects allow that the overall dust content, or covering factors, may be intrinsically different, however, such that the most obscured sources have larger covering factors and are not systematically obscured only because of orientation \citep{lev07,tho09,eli12}.  

The comparison of our results between Figure 3 and Figure 4 show obvious differences in the SEDs of unobscured quasars compared to obscured quasars.  Compared to unobscured sources in Figure 3, the obscured sources in Figure 4 have more far infrared luminosity (at $\sim$ 100 \ums) relative to the mid infrared 7.8 \um luminosity where the SEDs are normalized.  This is true also for the lower luminosity AGN. How do we interpret this result?  

Consider first the difference in $L_{\nu}$(100 \ums)/$L_{\nu}$(7.8 \ums) between unobscured quasars (Figure 3) and obscured quasars (Figure 4). The median ratio log $L_{\nu}$(100 \ums)/$L_{\nu}$(7.8 \ums) for unobscured quasars is 0.6 but is 1.05 for obscured quasars, a difference of about 2 $\sigma$ compared to the dispersions within the ratios for each class.  This means the unobscured quasars appear to have a smaller fraction of cool dust (seen in the far infrared) compared to hot dust; if this difference is intrinsic rather than an orientation effect, it means that there are real differences in the dust distribution between obscured and unobscured quasars.  The observed differences could be explained as arising only from orientation, however, if obscuration is so great that obscured quasars suffer extinction of the continuum at 7.8 \um compared to 100 \um.    
  
The summary by \citet{dra89} shows that the extinction at 7.8 \um from silicate absorption alone is about 20\% of extinction at the peak of 9.7 \um silicate absorption, as measured in magnitudes.  The average spectrum of the obscured quasars used for our sample (Figure 2) has a silicate feature that absorbs about 50\% of the continuum, corresponding to extinction of 0.75 mag at peak extinction.  This implies extinction of $\sim$ 0.15 magnitude at 7.8 \ums. The difference in log $L_{\nu}$(100 \ums)/$L_{\nu}$(7.8 \ums) between unobscured and obscured quasars in Figures 3 and 4 is 0.5, or 1.3 magnitudes, which is much larger that the estimated extinction of 0.15 mag.  

This result implies that extinction effects arising from orientation do not explain the differences between unobscured and obscured.  However, silicates are not the only source of extinction for the continuum.  The prototype highly absorbed AGN is IRAS F00183-7111 for which \citet{spo04} illustrate various other absorption features near 7.8 \um.  The features are normalized to the observed local continuum at 7.8 \um without any estimates of the actual extinction at 7.8 \ums, so it is feasible that these absorptions from ices and hydrocarbons suppress the 7.8 \um continuum by the additional $\sim$ one mag. needed to explain the differences in $L_{\nu}$(100 \ums)/$L_{\nu}$(7.8 \ums) ratios. In this case, orientation effects alone could explain why obscured quasars appear to have relatively higher far infrared luminosities. In sum, we cannot confidently conclude as yet that the differences in SEDs between unobscured and obscured quasars are caused by any effect other than differential extinction at 7.8 \um compared to 100 \ums, because of the complex absorptions near 7.8 \um.

\begin{figure}

\figurenum{3}
\includegraphics[scale= 1.0]{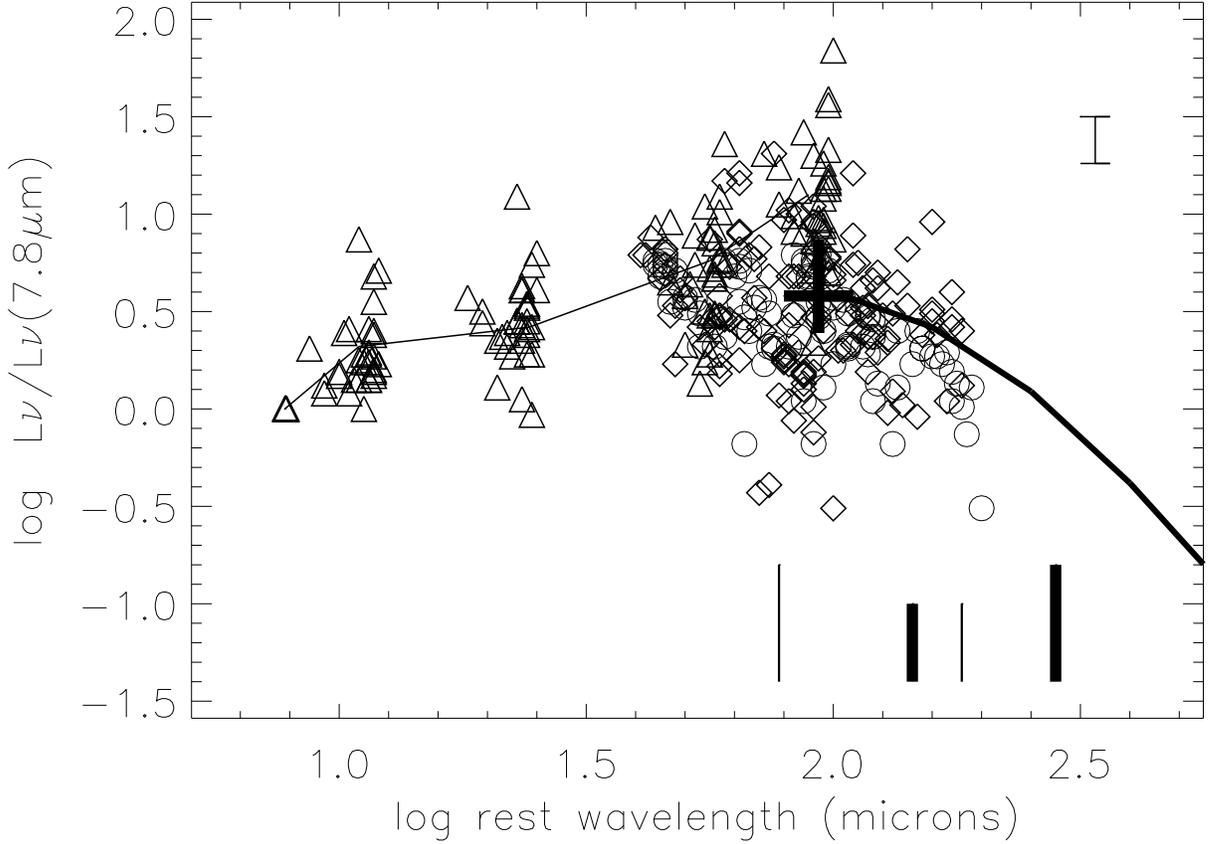}
\caption{Observed results for $L_{\nu}$/$L_{\nu}$(7.8 \ums) for unobscured AGN and quasars. Triangles are low redshift silicate emission AGN in \citet{sar11} with far infrared luminosities from IRAS photometry and $L_{\nu}$(7.8 \ums) from IRS spectra. Diamonds are high redshift SDSS/WISE quasars with new SPIRE photometry in Table 2, and circles are these high redshift quasars with upper limits. Large cross is the median and one sigma dispersion within rest wavelength range 80 \um to 110 \um for the high redshift quasars, including limits. Thin curve is the median for silicate emission AGN; thick curve is the most luminous ULIRG SED from Herschel photometry in \citet{sym13}, normalized at 100 \um to the observed median of the SDSS/WISE quasars.  Long, thick vertical line is the rest wavelength for source with z = 2.1 at observed frame 850 \um (SCUBA-2); short, thick vertical line is the rest wavelength for source with z = 2.1 at observed frame 450 \um (SCUBA-2). Long, thin vertical line is the rest wavelength for source with z = 10 at observed frame 850 \ums; short, thin vertical line is rest wavelength for source with z = 10 at observed frame 2 mm (GISMO). Luminosity ratios at these rest wavelengths are taken from thick curve. Error bar is observational uncertainty in ratio log $L_{\nu}$/$L_{\nu}$(7.8 \ums) for individual SDSS/WISE quasars. 
}

\end{figure}

\begin{figure}

\figurenum{4}
\includegraphics[scale= 1.0]{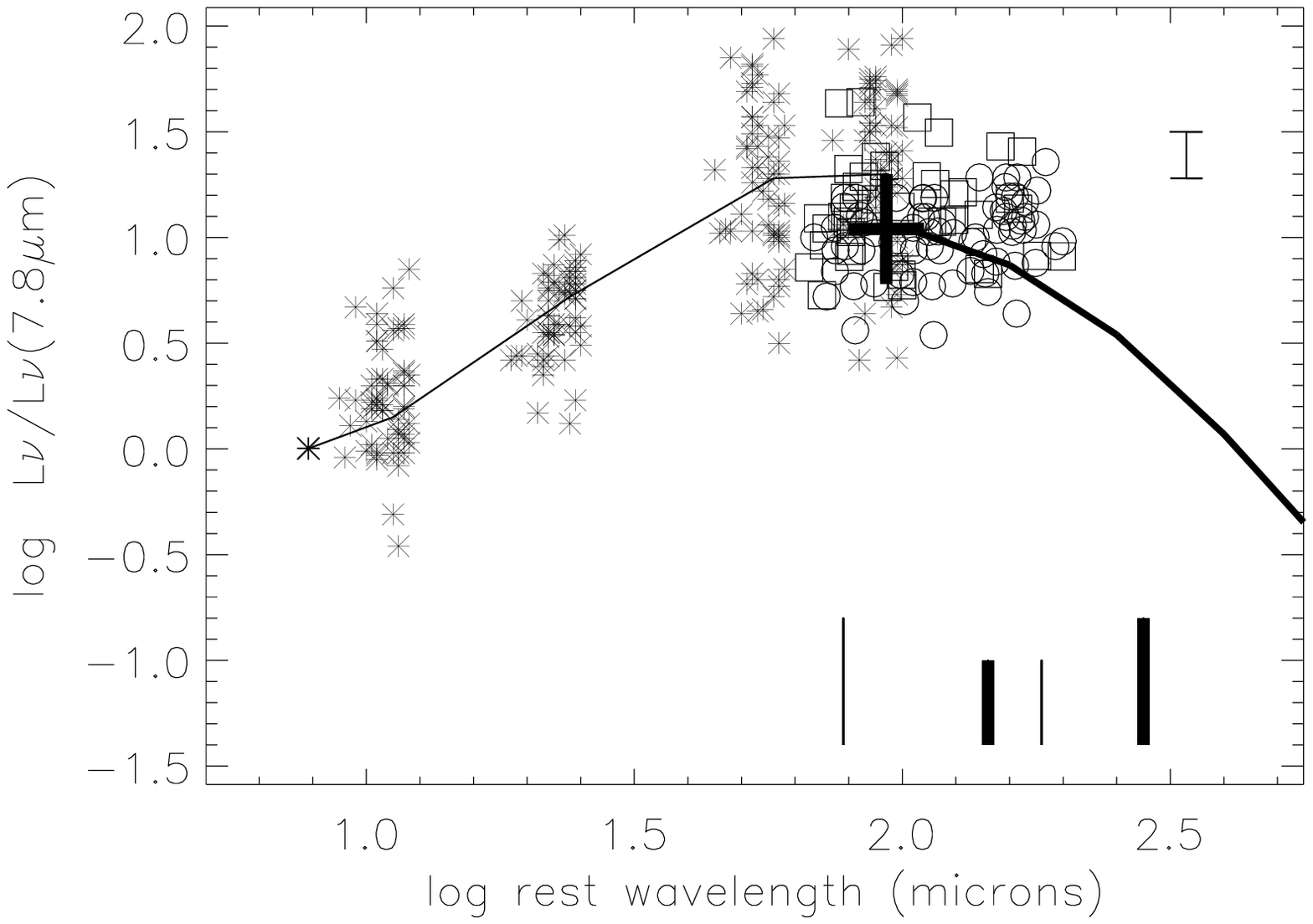}
\caption{Observed results for $L_{\nu}Lv(100 um)/Lv(7.8 um) $/$L_{\nu}$(7.8 \ums) for obscured AGN and quasars. Asterisks are low redshift silicate absorption AGN in \citet{sar11} with far infrared luminosities from IRAS photometry and $L_{\nu}$(7.8 \ums) from IRS spectra. Squares are high redshift obscured quasars discovered by $Spitzer$ IRS in Table 1, and circles are quasars with limits. Large cross is the median and one sigma dispersion within rest wavelength range 80 \um to 110 \um for the high redshift quasars, including limits. Thin curve is the median for silicate absorption AGN; thick curve is the most luminous ULIRG SED from Herschel photometry in \citet{sym13}, normalized at 100 \um to the observed median of the high redshift quasars.  Long, thick vertical line is the rest wavelength for source with z = 2.1 at observed frame 850 \um (SCUBA-2); short, thick vertical line is the rest wavelength for source with z = 2.1 at observed frame 450 \um (SCUBA-2). Long, thin vertical line is the rest wavelength for source with z = 10 at observed frame 850 \ums; short, thin vertical line is rest wavelength for source with z = 10 at observed frame 2 mm (GISMO). Luminosity ratios at these rest wavelengths are taken from thick curve. Error bar is observational uncertainty in ratio log $L_{\nu}$/$L_{\nu}$(7.8 \ums) for individual quasars. 
}

\end{figure}

There also are differences in both unobscured and obscured samples in log $L_{\nu}$(100 \ums)/$L_{\nu}$(7.8 \ums) for luminous quasars compared to local AGN.  Figure 3 shows that the high luminosity, unobscured quasars have log $L_{\nu}$(100 \ums)/$L_{\nu}$(7.8 \ums) that is smaller by 0.5 than the median ratio for lower luminosity AGN.  For the obscured quasars and AGN, the difference between high luminosity quasars and lower luminosity AGN is 0.25.  These differences can be attributed primarily to selection effects, because the quasars in both samples were selected based on brightnesses near rest frame 7.8 \ums, so their selection favors sources having smaller $L_{\nu}$(100 \ums)/$L_{\nu}$(7.8 \ums) compared to local AGN whose selection was not biased by 7.8 \um luminosities.  

An alternative interpretation of the systematic differences between obscured and unobscured quasar samples might invoke luminosity dependence in the $L_{\nu}$(100 \ums)/$L_{\nu}$(7.8 \ums) ratio, because the SDSS/WISE unobscured quasars are systematically more luminous by a factor of $\sim$ 10 than the Bo{\"o}tes obscured quasars (Figure 1).  We rule out this interpretation because it is not evident among the AGN.  These have similar luminosities between obscuration categories (Figure 1), but comparison of Figures 3 and 4 shows that the $L_{\nu}$(100 \ums)/$L_{\nu}$(7.8 \ums) ratio is a factor of 2.2 larger for the obscured AGN compared to unobscured.  This is similar to the factor of 2.8 for the difference between obscured and unobscured quasars at much higher luminosities. 

The dispersions in the $L_{\nu}$(100 \ums)/$L_{\nu}$(7.8 \ums) ratios in Figures 3 and 4 are a measure of the intrinsic variations in the ratio of cool dust to hot dust within sources.  Such variations can arise for many reasons, but it is useful to measure the extent of these variations.  To estimate intrinsic dispersions, the dispersions produced by observational uncertainties in measures of both $f_{\nu}$(100 \ums) and $f_{\nu}$(7.8 \ums) need to be removed. 

For the unobscured SDSS/WISE quasars in Figure 3, the uncertainties noted in Table 2 include $\pm$ 15\% for WISE 22 \um fluxes and $\pm$ 25\% for typical SPIRE fluxes.  In addition, there is additional uncertainty of $\sim$ $\pm$ 15\% at typical redshifts in using the template that transforms observed frame $f_{\nu}$(22 \ums) to rest frame $f_{\nu}$(7.8 \ums), as described in \citet{var14}.  Adding these uncertainties quadratically leads to an overall dispersion expected from observational uncertainty alone of $\pm$ 33\%.  This is shown as the error bar in Figure 3 and compares to the observed dispersion which shows a one $\sigma$ range of a factor of 3, or $\pm$ 50\%, in the $L_{\nu}$(100 \ums)/$L_{\nu}$(7.8 \ums) ratio (cross in Figure 3). From these comparisons of observational uncertainties and observed dispersions, we conclude that the intrinsic variation in the ratio of cool dust to hot dust is $\sim$ $\pm$ 40\%.

For the obscured quasars in Figure 4, the only uncertainty entering the rest frame $f_{\nu}$(7.8 \ums) is the 10\% uncertainty in measurement of the IRS spectrum.  Combining this with the 25\% SPIRE uncertainty gives a total observational uncertainty in $L_{\nu}$(100 \ums)/$L_{\nu}$(7.8 \ums) of $\pm$ 28\%, which compares to the observed dispersion (cross in Figure 4) of $\pm$ 50\%.  This yields an intrinsic variation in cool dust to hot dust for obscured quasars also of $\sim$ $\pm$ 40\%. 

To make predictions for submm or mm observations that see rest frame wavelengths longer than 100 \ums, the observed SEDs in Figures 3 and 4 for luminous quasars need to be extended. We do this by adopting the SED of the most luminous ULIRGs determined by $Herschel$ (Figure 17 of Symeonidis et al. 2013) and normalizing to the 100 \um luminosity of the quasars.  These extended SEDs are shown as thick curves in Figures 3 and 4.  These curves are used below to determine values of $L_{\nu}$($\lambda$)/$L_{\nu}$(7.8 \ums) for $\lambda$ the rest frame wavelength corresponding to submm and mm observations at different observed wavelengths and redshifts.

\subsection{Bolometric Dust Luminosities $L_{IR}$ for Obscured and Unobscured Quasars}

Having full SEDs allows the determination of total infrared luminosities $L_{IR}$. For the local AGN, the total infrared luminosity $L_{IR}$ reradiated by absorbing dust can be determined as defined by \citet{san96} using IRAS fluxes, whereby  $f_{IR}$ =  1.8 x 10$^{-11}$[13.48$f_{\nu}$(12) + 5.16$f_{\nu}$(25) + 2.58$f_{\nu}$(60) + $f_{\nu}$(100)], for $f_{IR}$ in erg cm$^{-2}$ s$^{-1}$ and IRAS flux densities in Jy. (This relation also includes an estimated contribution from longer wavelengths.) For AGN, \citet{sar11} found that log [$L_{IR}$/$\nu L_{\nu}$(7.8 \ums)] = 0.51 $\pm$ 0.21 in AGN with silicate emission and log [$L_{IR}$/$\nu L_{\nu}$(7.8 \ums)] = 0.80 $\pm$ 0.25 in AGN with silicate absorption.  

\begin{figure}

\figurenum{5}
\includegraphics[scale= 1.0]{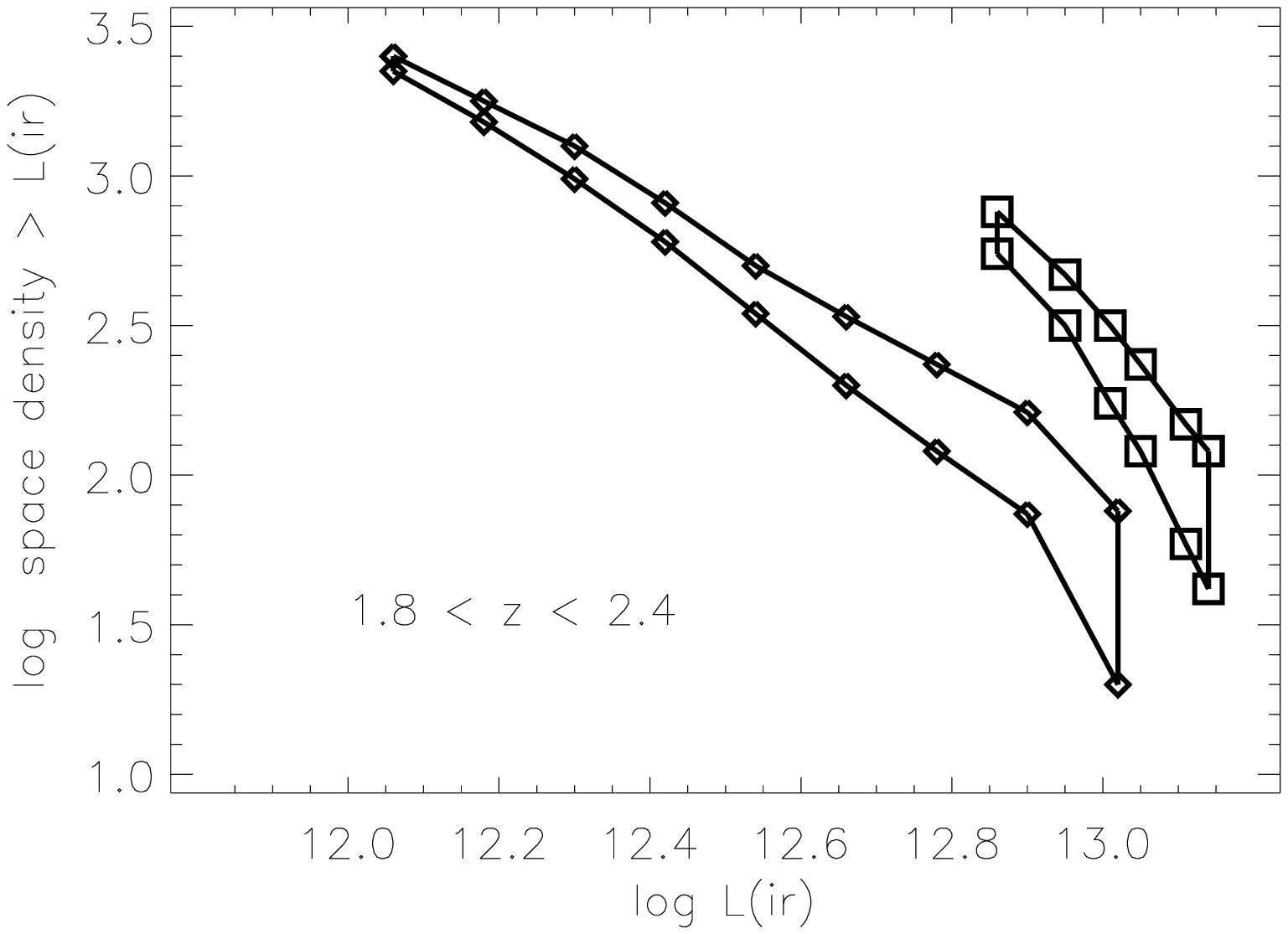}
\caption{Luminosity functions in $L_{IR}$ (\ldot) for unobscured quasars (diamonds) and obscured quasars (squares) with 1.8 $<$ z $<$ 2.4.   Space densities are number of quasars Gpc$^{-3}$ scaled from $\nu L_{\nu}$(7.8 \ums) space densities in \citet{var14} using relations log [$L_{IR}$/$\nu L_{\nu}$(7.8 \ums)] = 0.41 for unobscured quasars and log [$L_{IR}$/$\nu L_{\nu}$(7.8 \ums)] = 0.68 for obscured quasars, determined from Figures 3 and 4 as described in text.  The envelopes encompass statistical uncertainties $\pm$ $\sqrt{N}$ for N the number of quasars $>$ $L$ in this redshift interval and are shown only for luminosities that include quasars observed within the $\sim$ 8 deg$^{2}$ Bo{\"o}tes survey field; no extrapolations of luminosity functions to fainter sources than observed have been applied.}

\end{figure}

Using the results in Figures 3 and 4, these ratios can be modified by accommodating decreased relative flux densities at 60 \um and 100 \um for the high redshift quasars, assuming that shorter wavelengths which are unobserved retain the same ratios to $f_{\nu}$(7.8 \ums) and that $f_{\nu}$(60 \ums) = $f_{\nu}$(100 \ums) for the quasars. The result for the high redshift silicate emission quasars is log [$L_{IR}$/$\nu L_{\nu}$(7.8 \ums)] = 0.41 and for silicate absorption quasars is log [$L_{IR}$/$\nu L_{\nu}$(7.8 \ums)] = 0.68.  Applying these transformations to the $\nu L_{\nu}$(7.8 \ums) luminosity functions for obscured and unobscured quasars with 1.8 $<$ z $<$ 2.4 given in \citet{var14} 
gives the luminosity functions in Figure 5.  Although luminosity functions are similar in $\nu L_{\nu}$(7.8 \ums), the different corrections to $L_{IR}$ mean that the obscured quasars dominate the bolometric luminosity function.

One of the questions we are asking is what fraction of high redshift, high luminosity DOGS are powered by quasars, compared to the fraction powered by starbursts.  This is fundamental to deciding if luminous, obscured submm sources trace SFR in the early universe.  The $Herschel$ HerMES survey yielded an independent estimate of DOG luminosity functions \citep{cal13}.  We compare space densities of detected sources reported in this survey within the interval 1.5 $<$ z $<$ 2.5 with those that can be determined from our luminosity function in Figure 5.  We consider only the obscured quasars, because these are quasars which meet the DOG selection criteria.  Also, we have to choose luminosities bright enough that they overlap our luminosity function, so we can only use the brightest bin of the HerMES survey.  

In the 2 deg$^{2}$ of the HerMES survey, Calanog et al. report (their Table 2) 31 sources in this redshift interval having log $L_{IR}$ $>$ 12.85 \ldot, which yields a result of 1400 DOGS Gpc$^{-3}$ in this luminosity range. Transforming to the value we adopt for H$_0$ would be equivalent to log $L_{IR}$ $>$ 12.8 in Figure 5, above which luminosity are $\sim$ 800 obscured quasars Gpc$^{-3}$.  Their $L_{IR}$ for these sources are derived in different manner than ours by assuming various spectral templates so results for the luminosity functions are independent.  Given the uncertainties entering this comparison, these space densities are similar, which indicates that for the most luminous DOGS, the high redshift examples are dominated by DOG quasars rather than by DOG starbursts.  This result cautions against using DOG samples having no spectral classification as indicators of SFR. The dominance of quasars in the HerMES DOG study probably arises because they are found using a 24 \um criterion, which selects in favor of hotter dust. 

\section{Quasar Counts based on Dust Luminosities}

Our objective in this section is to compare source counts, observed and predicted, for all three UV/IR categories of dusty quasars within different redshift ranges and at different observing wavelengths, from mid-infrared to millimeter.  Because our SEDs scale to $L_{\nu}$(7.8 \ums), source counts are first established using observed $f_{\nu}$ at rest wavelength 7.8 \ums.  Source counts at submillimeter and millimeter wavelengths are then predicted by scaling the far infrared SEDs from section 3.2.  

\subsection {Quasar Counts for 1.8 $<$ z $<$ 2.4}

Obscured quasars discovered with the IRS based on the 9.7 \um silicate absorption feature are found at redshifts 1.5 $\la$ z $\la$ 3.5, and the redshift distribution of sources in Table 1 is shown in Figure 1.  The selection effect of having the 7.8 \um peak within the 24 \um $Spitzer$ MIPS survey band causes the sample to cluster within 1.8 $<$ z $<$ 2.4.  This is the redshift interval for which we have the best empirical result for counts of obscured quasars. 
The obscured quasars with IRS redshifts were found in spectroscopic follow up observations of sources discovered in FLS and Bo{\"o}tes, and Table 1 includes all sources in these survey fields with IRS spectra having $f_{\nu}$(24 \ums) $>$ 1 mJy and $R$ $>$ 24.  Our goal is to determine source counts for obscured quasars meeting these two criteria.  Because IRS spectra were not obtained of all sources defined by these flux limits, corrections need to be determined for incompleteness in the spectroscopic selections.  We determine these corrections by the ratio of sources having IRS spectra compared to the total number of sources meeting the photometric criteria.  

\begin{figure}

\figurenum{6}
\includegraphics[scale= 1.0]{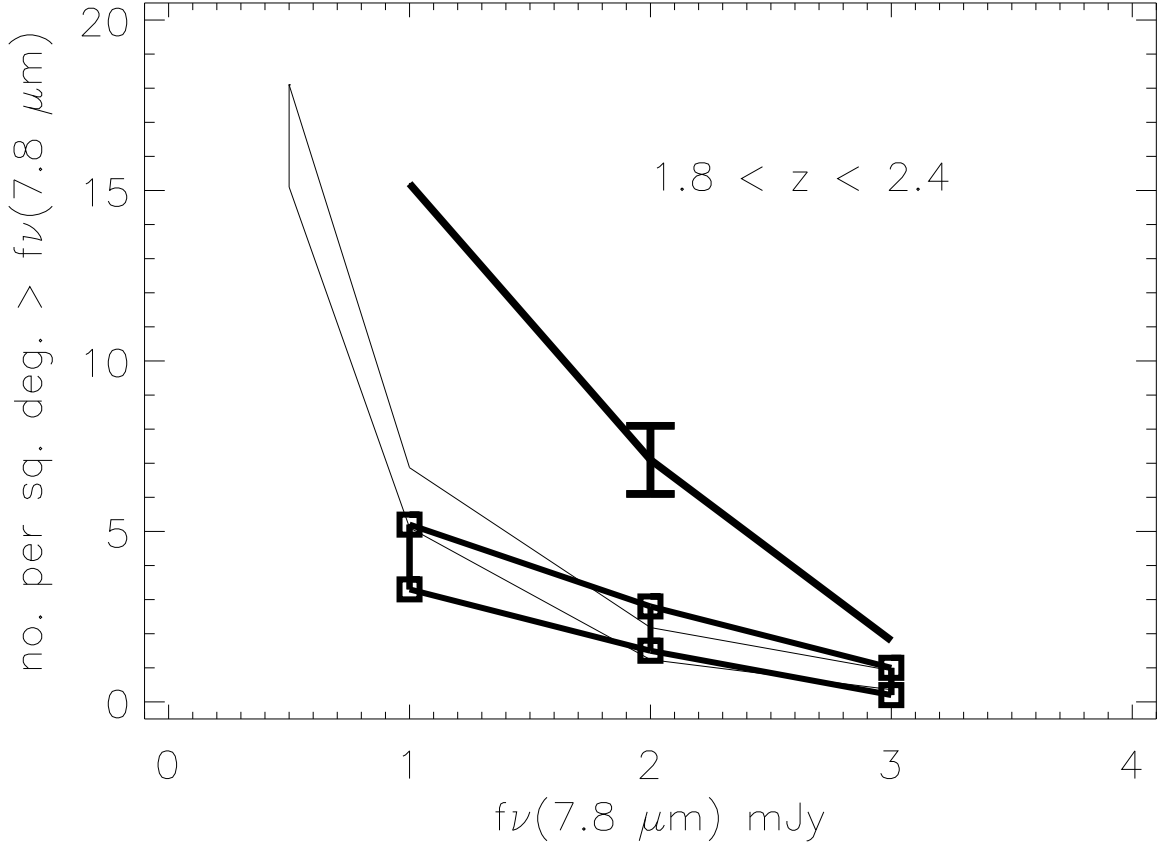}
\caption{Observed quasar counts within 1.8 $<$ z $<$ 2.4 for $f_{\nu}$ at rest wavelength 7.8 \um for unobscured and obscured quasars, corrected for incompleteness as described in text.  Envelope with thin line shows unobscured AGES quasars in Bo{\"o}tes with optical redshifts. Error bars with squares and envelope with thick line are obscured quasars in Bo{\"o}tes with IRS redshifts from silicate absorption having $R$ $>$ 24, listed in Table 1. Lengths of error bars and sizes of envelopes show statistical uncertainties $\pm$ $\sqrt{N}$ for N the total number of observed quasars $>$ $f_{\nu}$(7.8 \ums). Counts are shown only for quasars observed within the Bo{\"o}tes survey field; no extrapolations of counts to brighter or fainter sources have been applied. Source counts at any other observed wavelength or redshift can be predicted by scaling $f_{\nu}$($\lambda$)/$f_{\nu}$(7.8 \ums) from Figures 3 and 4 with $\lambda$ the rest wavelength corresponding to $\lambda_{observed}$/(1+z), as in following figures.  As discussed in text, intermediate quasars that are partially obscured are estimated as equal in number to the obscured quasars shown.  Total counts for all quasars is the sum of all three samples, shown as the single thick line, with statistical uncertainty shown by the error bar. } 

\end{figure}

Selection criteria varied between the FLS and Bo{\"o}tes spectroscopic surveys.  For the FLS, a variety of photometric criteria were used including IRAC colors and optical $R$ mag as bright as 19 \citep{saj07} whereas the Bo{\"o}tes spectroscopy used only 24 \um and optical criteria because the primary goal was to understand the optically faintest sources.  As a result, the Bo{\"o}tes sources contain many more obscured quasars despite the smaller overall spectroscopic sample size.  From the FLS survey, there are 11 obscured quasars in Table 1 within 1.8 $<$ z $<$ 2.4, but there are 22 from Bo{\"o}tes.  In addition, the optical component of the Bo{\"o}tes survey (NDWFS) reaches more than a magnitude fainter than the FLS survey, so faint magnitudes defining $R$ $\ga$ 24 are more reliable.  For these reasons, we use only the Bo{\"o}tes obscured quasars to determine incompleteness corrections and statistical uncertainties for the obscured quasars. 

The distribution of the Bo{\"o}tes survey in $R$ and [24 \ums] Vega magnitude is illustrated in Figure 1 of \citet{dey08}.  There are 85 sources having $R$ $>$ 24 and $f_{\nu}$(24 \ums) $>$ 1 mJy within an area of 8.2 deg$^{2}$, of which 53 are included in the spectroscopic samples summarized in \citet{hou05}, \citet{wee06}, and \citet{bus09}. This gives a correction of 1.6 for incompleteness.  (This is similar to the factor of 1.8 previously reported by \citet{wee06} as the ratio of total/observed Bo{\"o}tes sources based on a selection criterion of $f_{\nu}$(24 \ums) $>$ 1 mJy and $I$ $>$ 24.) After applying these correction factors and adding statistical uncertainties of $\pm$ N$^{0.5}$ for N the number of sources in a bin, the surface densities of obscured Bo{\"o}tes quasars with 1.8 $<$ z $<$ 2.4 are shown in Figure 6.

An additional advantage of using only Bo{\"o}tes for obscured quasars is that the AGES redshifts of unobscured quasars \citep{koc12} arise from the same Bo{\"o}tes survey area, although slightly smaller at 7.7 deg$^{2}$.  AGES is also based on an infrared selection criteria, $f_{\nu}$(24 \ums) $>$ 0.3 mJy, so that counts based on rest frame 7.8 \um flux densities arising from dust luminosities can be compared directly to the obscured quasar counts.  For quasars, AGES reaches $I$ $<$ 22.5.  The main survey fields within 7.7 deg$^{2}$ contain 2070 MIPS-selected quasars for which optical spectra were obtained of 1991 (Table 3 of Kochanek et al.) for a spectroscopic survey completeness of 96\%. Almost all are type 1 quasars; Figure 7 of \citet{hic07} shows that less than 2\% of sources with spectra are type 2.  Assuming that 98\% of sources are type 1 with a completeness correction of 1.04 results in an overall correction to counts for unobscured, type 1 quasars of only 1.02 times number of sources with spectra.  Taking the quasars from the AGES catalog having 1.8 $<$ z $<$ 2.4 and combining with our adopted infrared template results in the surface densities shown in Figure 6. (Uncertainties in the adopted template have little effect on the results because the observed frame 24 \um is close to rest frame 7.8 \um at the redshifts of interest.)

The results in Figure 6 show that the obscured and unobscured Bo{\"o}tes quasars are very similar in number for 1.8 $<$ z $<$ 2.4, as shown previously by \citet{var14}.  A similar conclusion was reached using a separate sample of obscured quasars - the "Extremely Luminous Infrared Galaxies" (ELIRGS).  These are the most infrared luminous quasars chosen with color selection criteria in the all sky WISE survey \citep{eis12,tsa15} and are interpreted as hot, dust obscured quasars \citep{asf15}.  The 20 sources tabulated in Tsai et al. are typically at z $\sim$ 3 and have median $f_{\nu}$(22 \ums) $\sim$ 13 mJy, or $f_{\nu}$(rest frame 7.8 \ums) $\sim$ 17 mJy with our assumed template. The median $R$ band flux densities are $\sim$ 2 $\mu$Jy (about magnitude 23), which gives representative log UV/IR $\sim$ -2.4, meeting our definition of obscured quasars.  Similar space densities for these obscured ELIRGS compared to unobscured quasars were determined by \citet{asf15} at the highest luminosities within 2.0 $<$ z $<$ 2.5 by comparing the ELIRGS to SDSS quasars.   

Determining accurate counts for the partially obscured quasars (primarily type 2) having -1.8 $<$ log UV/IR $<$ 0.2 is more uncertain because they are not well represented in either the IRS or AGES spectra.  We estimate the numbers for partially obscured quasars in two alternative ways.  First, we note that the photometric redshift estimates in Table 3 of \citet{hic07} for AGES quasars identified by X-ray criteria indicate 45 IRAGN1 and 76 IRAGN2 within 2 $<$ z $<$ 2.5.  This indicates that counts of partially obscured, type 2 quasars exceed unobscured within our redshift interval by $\sim$ 1.7.  The numbers of type 2 quasars at high redshifts are particularly uncertain, however, because of the absence of spectroscopic redshifts \citep{bro06}. 

The second estimate derives from the distribution of optical magnitudes for all quasars showing silicate absorption in the FLS spectral surveys \citep{yan07,saj07,das09,wee06b}.  For this purpose, the FLS is useful rather than Bo{\"o}tes because the FLS surveys were not restricted by optical magnitudes, but any quasar showing silicate absorption in the infrared spectrum must have some obscuration even if brighter than the limit of $R$ $>$ 24 for obscured quasars. From the FLS survey area, Table 1 includes 11 obscured sources with 1.8 $<$ z $<$ 2.4 and $R$ $>$ 24.  There are an additional 11 IRS observed quasars within 1.8 $<$ z $<$ 2.4 having silicate absorption (which we verified in CASSIS) and 22 $<$ $R$ $<$ 24 which are not included in Table 1 (MIPS numbers 8226, 268, 8251, 521, 509, 22204, 16080, 16152, 22482, 15949, and 16113.)  Although statistics are small, this equal number indicates that the sample of partially obscured quasars ($R$ $<$ 24) is the same as the obscured, optically faint sample ($R$ $>$ 24). 

These two comparisons of obscured, partially obscured and unobscured quasars lead to our estimate that partially obscured quasars are equal in number to either obscured or unobscured quasars.  This result is illustrated in Figure 6 for quasar counts at rest frame 7.8 \um. The sum of quasar counts shown in Figure 6 includes all quasars, therefore, with equal contributions from unobscured, partially obscured, and obscured quasars as defined by UV/IR.

\section {Detections with Submm and mm Observations}

The evolution of SFR in the universe is tracked primarily by the evolution of sources with cool dust \citep{cha01, dal02, asf10, elb11, war11}, invoking the assumption that dust luminosity at $\ga$ 100 \um arises completely from star formation.  We want to test this assumption by determining how many quasars contaminate the submm samples, based on the empirical determinations given above of far infrared luminosities and space densities for all categories of dusty quasars.  In what follows, we determine the expected submm counts for these quasar populations and compare with sources actually known from the SCUBA-2 surveys \citep{cha05,bar14,ros13, gea13}.

\subsection{Dusty Quasars and Submm Sources with 1.8 $<$ z $<$ 2.4}

Although previous analyses have concluded from optical spectral classifications, X-ray observations and SED considerations that the submm source surveys contain few AGN \citep{cha05, ale05}, obscured and partially unobscured quasars would be difficult to identify in these ways so could have been overlooked as contributing to submm source counts. This is emphasized, for example, by Alexander et al. who find that only $\sim$ 8\% of SCUBA sources have observed X-ray fluxes consistent with quasars but that the fraction can increase to $\sim$ 80\% if absorbed X-rays are assumed in Compton thick, dusty sources.  This ambiguity is our main reason for comparing expected counts of the known dusty quasar population to actual submm detections.  Combining the rest frame 7.8 \um counts shown in Figure 6 with the SEDs in Figures 3 and 4 allows predictions of quasar counts that should be observed in submm source counts within 1.8 $<$ z $<$ 2.4. The expected counts at the 850 \um and 450 \um wavelengths of SCUBA-2 are shown in Figures 7 and 8. 

\begin{figure}

\figurenum{7}
\includegraphics[scale= 1.0]{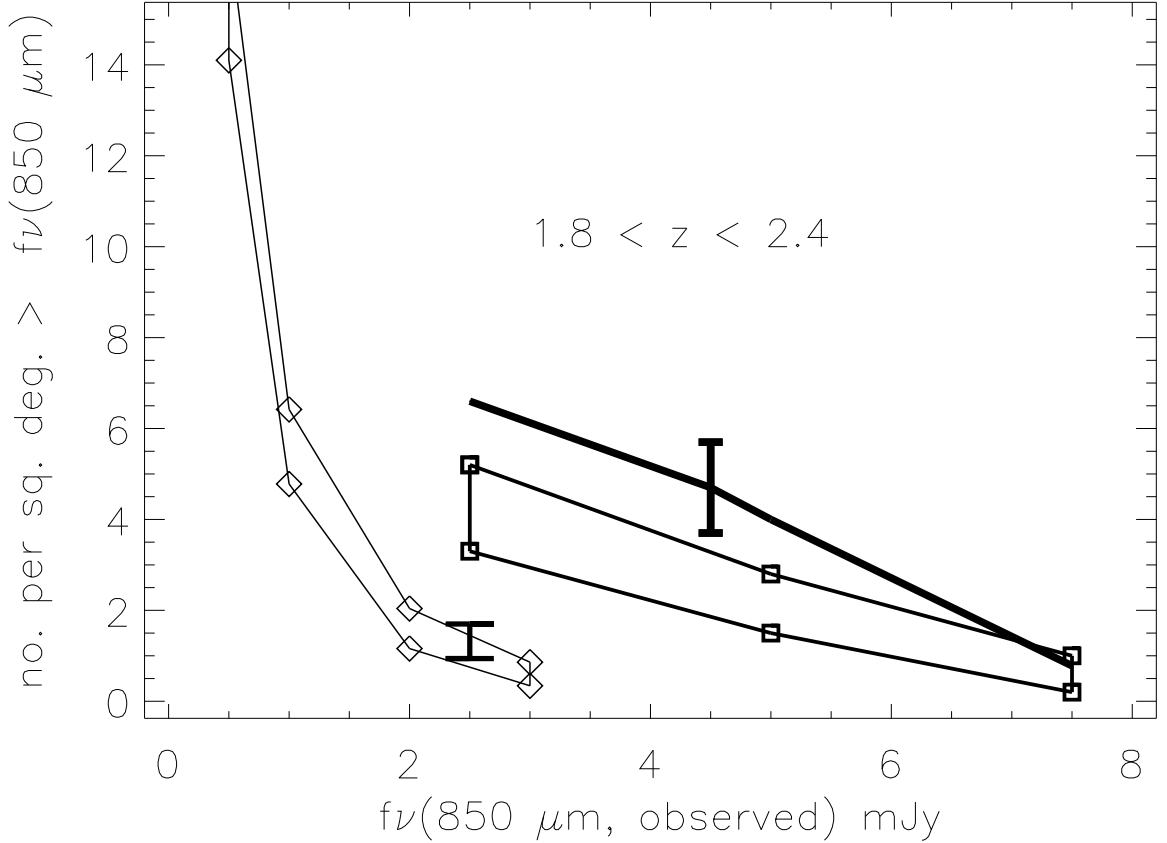}
\caption{Expected quasar counts within 1.8 $<$ z $<$ 2.4 at observed wavelength 850 \um (SCUBA-2) for quasars scaled from $f_{\nu}$(7.8 \ums) counts in Figure 6 using SEDs in Figures 3 and 4.  Diamonds are unobscured quasars scaled from AGES optical survey, and squares are optically faint, obscured quasars with IRS redshifts from silicate absorption. Range of counts encompasses statistical uncertainties in the source counts from Figure 6.  Small error bar shows estimate for the partially obscured quasars scaled as described in text, having 7.8 \um counts from Figure 6 the same as obscured quasars but assuming SEDs the same as unobscured quasars in Figure 3.  Total counts for all quasars is the sum of all three samples, shown as the single thick line, with statistical uncertainty shown by the thick error bar. Current SCUBA-2 850 \um detection limit is $\sim$ 2 mJy \citep{bar14}. } 

\end{figure}

\begin{figure}

\figurenum{8}
\includegraphics[scale= 1.0]{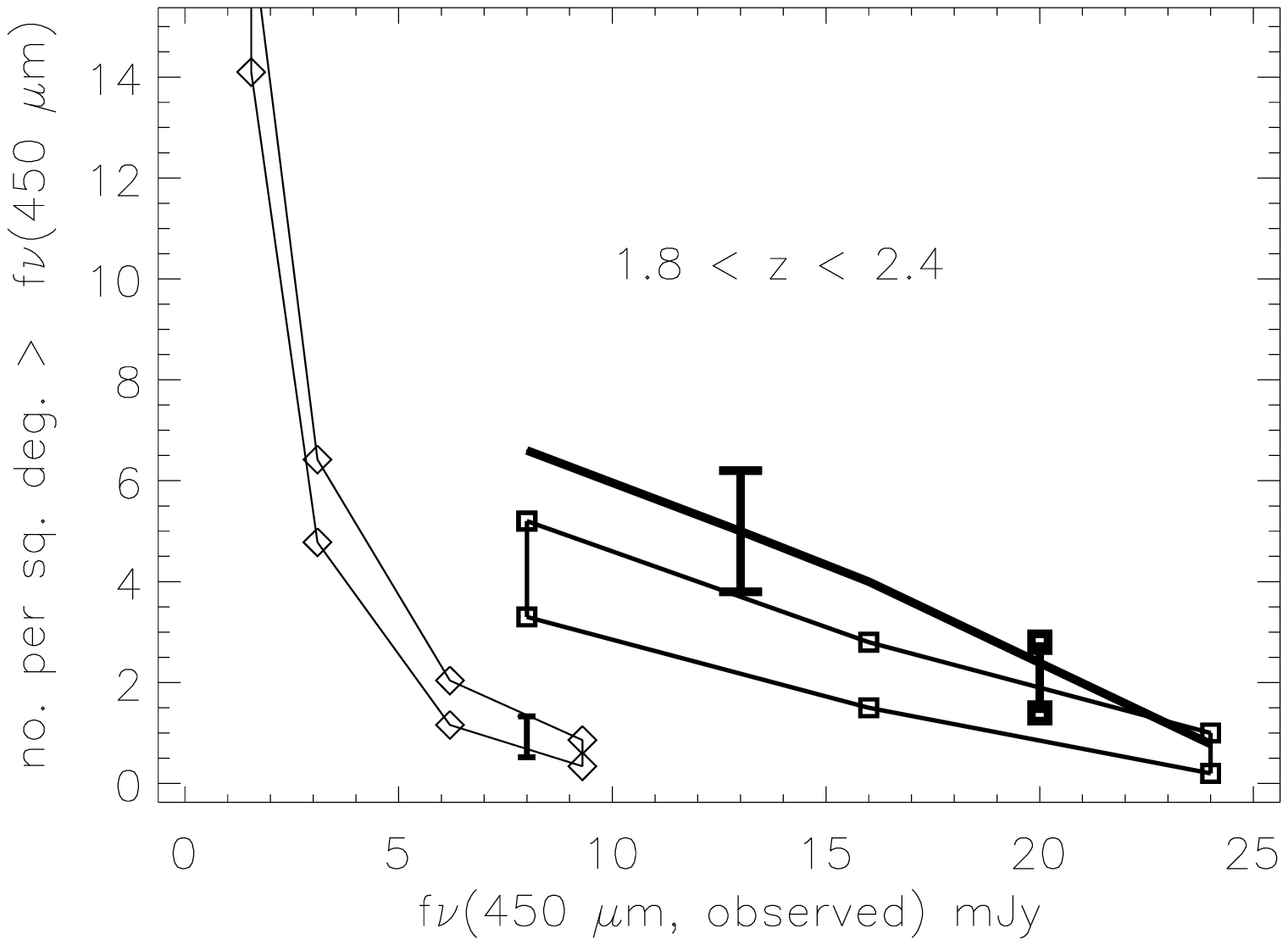}
\caption{Expected quasar counts within 1.8 $<$ z $<$ 2.4 at observed wavelength 450 \um (SCUBA-2) for quasars scaled from $f_{\nu}$(7.8 \ums) counts in Figure 6 using SEDs in Figures 3 and 4.  Diamonds are unobscured quasars scaled from AGES optical survey, and squares are optically faint, obscured quasars with IRS redshifts from silicate absorption.  Range of counts encompasses statistical uncertainties in the source counts from Figure 6.  Small error bar shows estimate for the partially obscured quasars scaled as described in text, having 7.8 \um counts from Figure 6 the same as obscured quasars but assuming SEDs the same as unobscured quasars in Figure 3. Total counts for all quasars is the sum of all three samples, shown as the single thick line, with statistical uncertainty shown by the thick error bar. Current SCUBA-2 450 \um detection limit is $\sim$ 6 mJy \citep{ros13}.  The vertical bar with squares at 20 mJy shows expected range of counts for obscured quasars at observed wavelength 350 \um and detection limit 20 mJy for comparison to SPIRE surveys such as in Table 1.} 

\end{figure}

From these Figures, it is seen that the expected submm counts from quasars are dominated by the obscured quasars, which are the sources which would not have been identified among spectroscopic redshifts of submm sources. This emphasizes why quantitative comparisons of expected quasars with observed counts are important.  For specific comparisons with submm surveys, we use the \citet{bar14} survey at 850 \um with SCUBA-2 and the \citet{ros13} 450 \um survey with SCUBA-2.  

The expected counts at 850 \ums, determined with our empirical 7.8 \um source counts in Bo{\"o}tes with no extrapolations, nearly reach the 2 mJy limit of the faintest 850 \um survey, GOODS-N in \citet{bar14}. The predicted counts for all quasars are $\sim$ 7 deg$^{-2}$ $>$ 2 mJy within 1.8 $<$ z $<$ 2.4 at observed frame 850 \um.  Barger et al. find five 850 \um sources with spectroscopic 1.8 $<$ z $<$ 2.4 in 400 arcmin$^{2}$ brighter than 2 mJy, or a density of 45 deg$^{-2}$.  If estimated photometric redshifts are added, there are 3 more sources for a total density of 72 deg$^{-2}$, ten times more than the expected number of quasars.  These submm detections are actually lower limits because the flux density limit is somewhat brighter over parts of the field.  

Of course, these results suffer from small number statistics, but they certainly confirm that the 850 \um surveys are indeed dominated by starbursts, as previously concluded by others. The infrared classification of quasars and starbursts based on the strength of PAH emission also confirms the dominance of starbursts in submm samples.  Of the submm sources observed with the $Spitzer$ IRS by \citet{pop08} and \citet{men09}, at least 80\% show PAH features.  The most useful result of our analysis is that the heavily obscured quasar population, not known before $Spitzer$, is not a significant contaminant for the 850 \um surveys. The main difference between this result and our conclusion in section 3.3 that high redshift DOGS detected by $Herschel$ SPIRE are dominated by quasars probably arises because the DOG selection is based on 24 \ums, which selects for the hotter dust of quasars.

Comparison to the SCUBA-2 450 \um survey \citep{ros13} gives even larger differences between observed counts and quasar counts, although the 450 \um redshifts are only photometric.  The 450 \um detection limit is 6 mJy, to which 19 separate sources were found within 210 arcmin$^{2}$ having photometric redshifts 1.8 $<$ z $<$ 2.4, for a density of 325 deg$^{-2}$ in this redshift interval.  From Figure 8, we would expect only $\sim$ 7 quasars deg$^{-2}$ to this limit. The large difference in densities of 450 \um compared to 850 \um sources is puzzling and seems to arise in part because of photometric redshift estimates, and in part because of a large variance in overall submm source densities between the survey fields.  

For example, 19/69 of the 450 \um sources in Roseboom et al. are assigned redshifts 1.8 $<$ z $<$ 2.4, which is consistent with the 24/73 of 850 \um sources with spectroscopic redshifts in \citet{cha05}.  However, only 8/49 of the 850 \um sources in \citet{bar14} have spectroscopic or photometric redshifts in this interval, with 23 sources assigned no redshift.  If these no redshift sources have comparable fractions within 1.8 $<$ z $<$ 2.4 as for the Chapman et al. 850 \um sources, this indicates that the Barger et al. results underestimate by about a factor of two the real number of sources within 1.8 $<$ z $<$ 2.4, which would raise the estimate to $\sim$ 150 deg$^{-2}$, about 1/2 the estimate from the 450 \um survey.     

Total counts at 850 \um and 450 \um also differ by about this same factor.  For any redshifts $\la$ 3, the ULIRG curve in Figures 3 and 4 shows that observed frame 450 \um observations should see flux densities about 3 times brighter for the same ULIRGS seen in observed frame 850 \um observations. Yet, the 450 \um survey reports 69 sources $>$ 6 mJy in 210 arcmin$^{2}$, or 1200 deg$^{-2}$, compared to the 850 \um result of 49 sources $>$ 2 mJy in 400 arcmin$^{2}$, or 440 deg$^{-2}$, so the surface density of 450 \um sources is nearly 3 times larger. This implies either a large incompleteness in the 850 \um results to the assumed 2 mJy limit, or a large cosmic variance in the survey fields.

Regardless of the explanation of differences between 850 \um and 450 \um surveys, we can conclude that the quasars in our known populations are responsible for only between 2\% and 10\% of known submm sources with 1.8 $<$ z $<$ 2.4.  The precise fraction may depend on observing wavelength and cosmic variance. 

We also illustrate in Figure 8 the detections that would be expected by SPIRE at 350 \ums.  The SPIRE surveys for sources with 1.8 $<$ z $<$ 2.4 should be most sensitive at this wavelength where rest frame wavelengths are closest to the SED maximum.  Scaling $f_{\nu}$(7.8 \um) to SPIRE rest frame wavelengths using the SEDs in Figures 3 and 4 and combining with counts from Figure 6 gives the result in Figure 8. The SPIRE detection limit is taken as 20 mJy by comparison to Bo{\"o}tes sources observed in Table 1.  The predicted number of detections of $\sim$ 2 deg$^{-2}$ for absorbed quasars compares well with the 12 obscured quasars in Table 1 detected in Bo{\"o}tes.  The result in Figure 8 also indicates that we would not expect any SPIRE detections of unobscured quasars (counts of unobscured quasars extrapolate to densities approaching zero at the 20 mJy limit required) so we predict that no AGES quasars in Bo{\"o}tes are detected by SPIRE.

\begin{figure}

\figurenum{9}
\includegraphics[scale= 1.0]{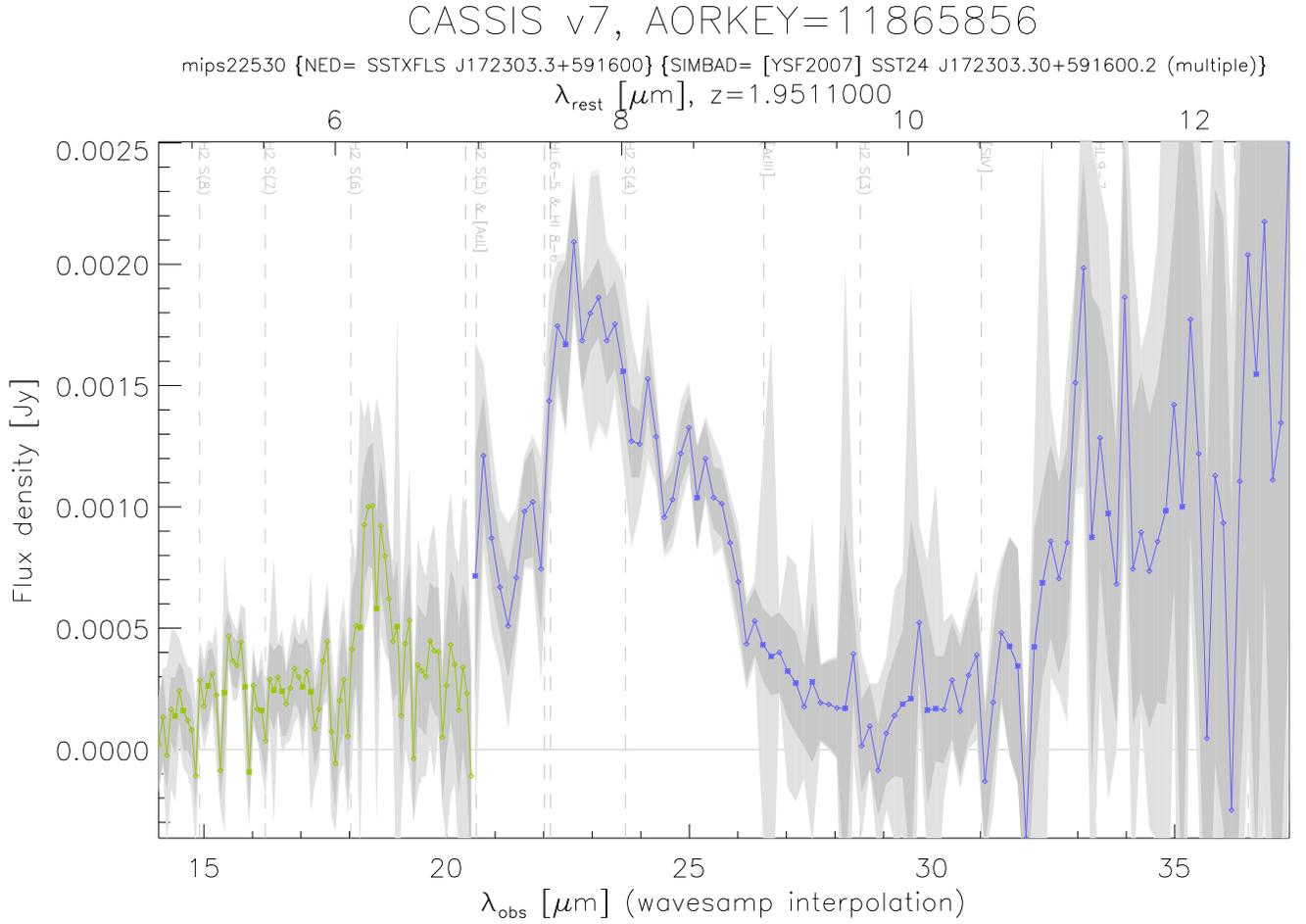}
\caption{Downloaded CASSIS spectrum for source MIPS22530 from FLS survey. Solid curve with points is the optimal spectrum determined by CASSIS.  Shading indicates uncertainties within individual spectral pixels.  Rest frame wavelengths at adopted redshift shown at top. The PAH 7.7 \um feature dominates the spectral flux, and a distinct 6.2 \um feature is also seen which can be used for starburst classification.} 

\end{figure}

\subsection{Comparing Submm Sources and $Spitzer$ Starbursts}

This large excess of submm sources compared to dusty quasars initially seems surprising because the number of high redshift obscured quasars is comparable to the number of high redshift PAH sources (classified as starbursts) in the $Spitzer$ IRS surveys.  We examine in more detail, therefore, whether the submm surveys based on detecting the rest frame far infrared continuum reveal the same starburst population as the $Spitzer$ IRS surveys which detect PAH features.   We can test this only within a redshift interval similar to that used for the obscured quasars, because the $Spitzer$ photometric surveys at 24 \um that reveal starbursts heavily favor redshifts at which the strong 7.7 \um PAH feature is within the 24 \um band, similarly to the redshift selection for the DOGS peaking at 7.8 \ums.  For this comparison, the FLS spectral surveys are most useful instead of the Bo{\"o}tes surveys because the FLS surveys are not constrained to faint optical magnitudes, and high redshift starbursts are not necessarily DOGS that would be fainter than $R$ $\sim$ 24.  An example of such a source found within the FLS survey is illustrated in Figure 9.

To make this test, we reexamined all FLS sources included in the SPIRE measures by \citet{saj12} for which a PAH detection is mentioned.  Within the range 1.8 $<$ z $<$ 2.4, the CASSIS spectra verified PAH features in 11 sources (MIPS289, MIPS521, MIOPS16113, MIPS16227, MIPS22417, MIPS22482, MIPS22530, MIPS22548, MIPS22633, AOR12507648, and AOR12508672).  These are extremely luminous starbursts.  Analogous to our scaling of AGN luminosities to the spectral peak at 7.8 \um, we scale PAH luminosities to the peak at rest frame 7.7 \um, $\nu L_{\nu}$(7.7 \ums).  As measured in CASSIS, all of these sources have $f_{\nu}$(7.7 \um) $>$ 1.5 mJy.  For $f_{\nu}$(7.7 \um) of 1.5 mJy at z = 2.0, log $\nu L_{\nu}$(7.7 \ums) =  45.8 (erg s$^{-1}$) or 12.2 (\ldot). Using local starbursts with IRS spectra and IRAS fluxes, \citet{sar11} calibrate log $L_{IR}$/$\nu L_{\nu}$(7.7 \ums) = 0.74.  From \citet{ken98} calibrating star formation rate (SFR) to total luminosity, log (SFR) = log $L_{IR}$ - 9.76, for SFR in \mdot and $L_{IR}$ in \ldot.  Luminosities log $\nu L_{\nu}$(7.7 \ums) $>$  12.2 \ldot~ for the PAH feature correspond, therefore, to log $L_{IR}$ $>$ 12.9 \ldot, or SFR $>$ 1300 \mdot.  

The SPIRE 350 \um flux densities in Sajina et al. allow a measure of the far infrared luminosity of these PAH sources compared to the $f_{\nu}$(7.7 \ums) from CASSIS.  For these 11 sources, we find that log $f_{\nu}$(115 \um)/$f_{\nu}$(7.7 \um) = 1.2 $\pm$ 0.1, at a SPIRE rest frame wavelength of 115 \um for the average z of 2.03.  The flux density observed by SPIRE can be scaled to that observed by SCUBA-2 850 \um using the long wavelength SED for luminous ULIRGS shown in Figure 3 or 4.  The resulting SCUBA-2 flux density that should be observed at 850 \um (275 \um rest frame for z = 2.1) would be $\sim$ 6 mJy for a PAH $f_{\nu}$(7.7 \ums) of 1.5 mJy.  Using this scaling, the expected count of SCUBA-2 sources can be determined from the observed count of PAH sources in the FLS survey.  

To compare with SCUBA-2 surveys, the FLS survey has to be corrected for incompleteness.  This correction is given in \citet{das09} who state that the sample is 57\% complete (to limits including our relevant PAH flux densities) over an area of 2.8 deg$^{2}$. Applying this correction to the 11 sources detected gives a density of 7 $\pm$ 2 PAH sources deg$^{-2}$ within 1.8 $<$ z $<$ 2.4 having $f_{\nu}$(7.7 \ums, rest frame) $>$ 1.5 mJy. From the above $f_{\nu}$ ratios, this corresponds to $f_{\nu}$(275 \ums, rest frame) $>$ 6 mJy for SCUBA-2 observations at 850 \ums.  Only very few SCUBA-2 sources are so bright. In \citet{bar14}, there are only 2 sources above 6 mJy in 400 arcmin$^{2}$ within 1.8 $<$ z$<$ 2.4, which gives a density of 18 $\pm$ 13 deg$^{-2}$. (Uncertainties in these count densities are scaled by N$^{-0.5}$ for N the number of actual sources which were found.)  Within the large statistical uncertainties that arise because of the few sources detected in either FLS or SCUBA-2 surveys, the results for PAH sources overlap the results for submm sources.  This indicates that similar starbursts are detected at high redshift with these independent methods, but the uncertainties are too large for a definitive conclusion about whether precisely the same populations are detected.  The best route to a final test will be to observe SCUBA-2 flux densities for numerous PAH sources detected by IRS.  This can determine if similar SFR densities are measured with both techniques.  Nevertheless, the comparison of these results for $Spitzer$ PAH starbursts and SCUBA-2 submm starbursts confirms that a large excess of dusty starbursts compared to dusty quasars should be expected at high redshifts, as observed for the submm sources.

\subsection{Quasars with 9.5 $<$ z $<$ 10.5}

As quasar discoveries continue to higher and higher redshifts, understanding the existence of the supermassive black holes required to produce their luminosity becomes increasingly puzzling \citep[e.g.][]{vol12,fen13,tof14}.  Quasars found to the highest redshifts seen so far (5 $<$ z $\la$ 7) are also dusty \citep{ven15,wu15}. There is potential to push dusty quasar discoveries to extreme redshifts, z $\ga$ 10, because surveys are now beginning at 2 mm with GISMO \citep{sta14} that have sufficient sensitivity to detect dusty sources at such redshifts \citep{dwe14}, as the observing band moves closer to the rest frame peak of luminosity.  It was this observational breakthrough that initially stimulated our investigation of potential dusty quasar detections at very high redshifts. Although SCUBA-2 850 \um detections cannot reach the same luminosity limits at these redshifts, having the observing band close to the rest frame luminosity peak also makes SCUBA-2 competitive for mapping the very high redshift universe. 

Sources detectable with GISMO or SCUBA-2 represent the best opportunity to constrain the very high redshift quasar population because redshifts are so large that any optical discoveries ($<$ 1 \ums) are precluded by intergalactic absorption below the Lyman limit.  As Staguhn et al. and Dwek et al. emphasize, detections of z $>$ 6 sources are one of the primary motivations for GISMO.  Even order-of-magnitude observational constraints are useful to determine if luminosity evolution is present and if the predominance of luminous starbursts over luminous quasars by the large factor seen at lower redshifts continues.

The nature of the overall dusty population at such extreme redshifts must be eventually determined observationally, but results derived above for z $\sim$ 2 provide a benchmark comparison, if the same luminosity function continues to higher redshifts. For scaling with the SEDs based on $L_{\nu}$/$L_{\nu}$(7.8 \ums), Figure 10 shows the quasar source counts for 9.5 $<$ z $<$ 10.5 if observations could be made at rest frame 7.8 \um for these redshifts (not currently feasible).  In Figure 11, these counts are converted to observed frame 850 \um and 2 mm using the counts in Figure 10 and the SEDs in Figures 3 and 4. The results in Figure 11 indicate that if quasar and starburst luminosity functions continue unchanged to such high redshifts, a source at z $\sim$ 10 may already have been found within existing 2 mm GISMO or 850 \um SCUBA-2 surveys! 

At the GISMO survey limit of 0.5 mJy, Figure 11 indicates that $\ga$ 5 quasars deg$^{-2}$ are expected within 9.5 $<$ z $<$ 10.5 for unchanging luminosity functions.  Taking the result from section 5.1 that the total number of submm sources observed at z = 2.1 exceeds the expected number of quasars by $\sim$ 20 implies 100 sources deg$^{-2}$ that should be found within 9.5 $<$ z $<$ 10.5 if the mix of quasars and starbursts continues the same.  The GISMO survey already includes 0.01 deg$^{2}$ and contains 5 unidentified sources, so it is feasible based on these estimates that one of these could be at z $\sim$ 10.  

To date, the 850 \um SCUBA-2 surveys are even more promising. The deepest survey limit is 2 mJy, at which Figure 11 shows an expected quasar density (dominated by obscured quasars) of 2 deg$^{-2}$.  If this is scaled by the factor of 20 to include luminous starbursts, a submm source with 9.5 $<$ z $<$ 10.5 statistically should be found when SCUBA-2 surveys to 2 mJy cover 0.025 deg$^{2}$. The 850 \um survey field in \citet{bar14} already covers 0.1 deg$^{2}$, so at least one of the optically unidentified sources already found in that field should be at such a redshift if quasar and starburst luminosity functions have not changed from z $\sim$ 2.1.  

These results are intended to show only that quasars like those already known at z = 2 would be detectable at z = 10 with current observational techniques.  Of course, we expect that luminosity functions and the starburst/quasar mix changes between z = 2 and z = 10.  There is as yet no observational proof of these changes, however, and these illustrative calculations are meant to show the potential of submm and mm surveys for providing the answers.  These results show the imperative of getting redshifts for unidentified GISMO and SCUBA-2 sources and of increasing the survey areas.   With only slight increases in survey area, the absence of any source with 9.5 $<$ z $<$ 10.5 will provide strong constraints on how the starburst/quasar luminosity functions diminish between redshifts of 2 and 10.

\begin{figure}

\figurenum{10}
\includegraphics[scale= 1.0]{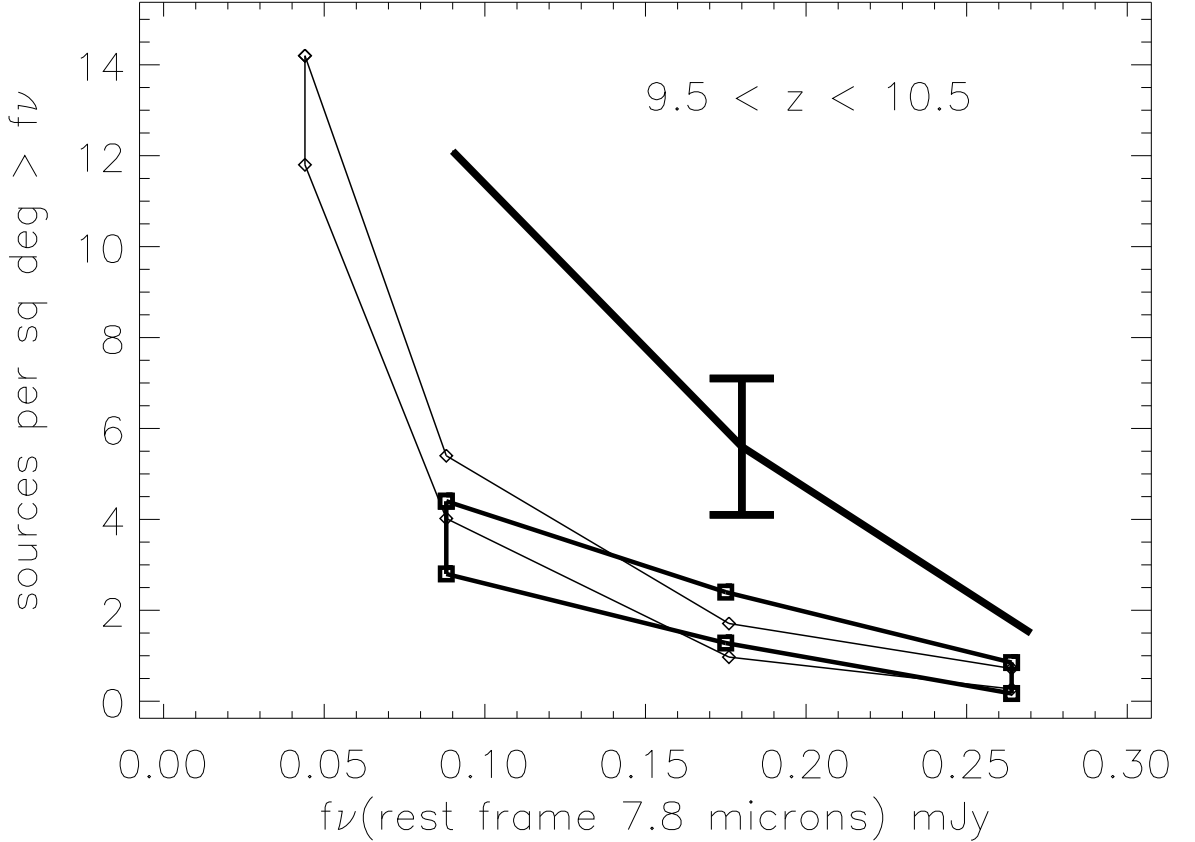}
\caption{Expected quasar counts within 9.5 $<$ z $<$ 10.5 at observed wavelengths corresponding to rest wavelength 7.8 \um for unobscured and obscured quasars if luminosity functions are the same for 9.5 $<$ z $<$ 10.5 as for 1.8 $<$ z $<$ 2.4.  Diamonds and thin envelope are unobscured quasars like those from AGES and squares with thick envelope are like obscured IRS quasars in Table 1.  Range of envelopes includes statistical uncertainties in the Bo{\"o}tes counts of these quasar populations.   Source counts at any observed wavelength can be predicted, as in Figures 7 and 8, by scaling $f_{\nu}$($\lambda$)/$f_{\nu}$(7.8 \ums) from Figures 3 and 4 with $\lambda$ the rest wavelength corresponding to $\lambda_{observed}$/(1+z). As discussed in text, intermediate quasars that are partially obscured are estimated as equal in number to the obscured quasars shown. Single thick line is sum for all three quasar classes with statistical uncertainty shown by error bar. }

\end{figure}

\begin{figure}

\figurenum{11}
\includegraphics[scale= 1.0]{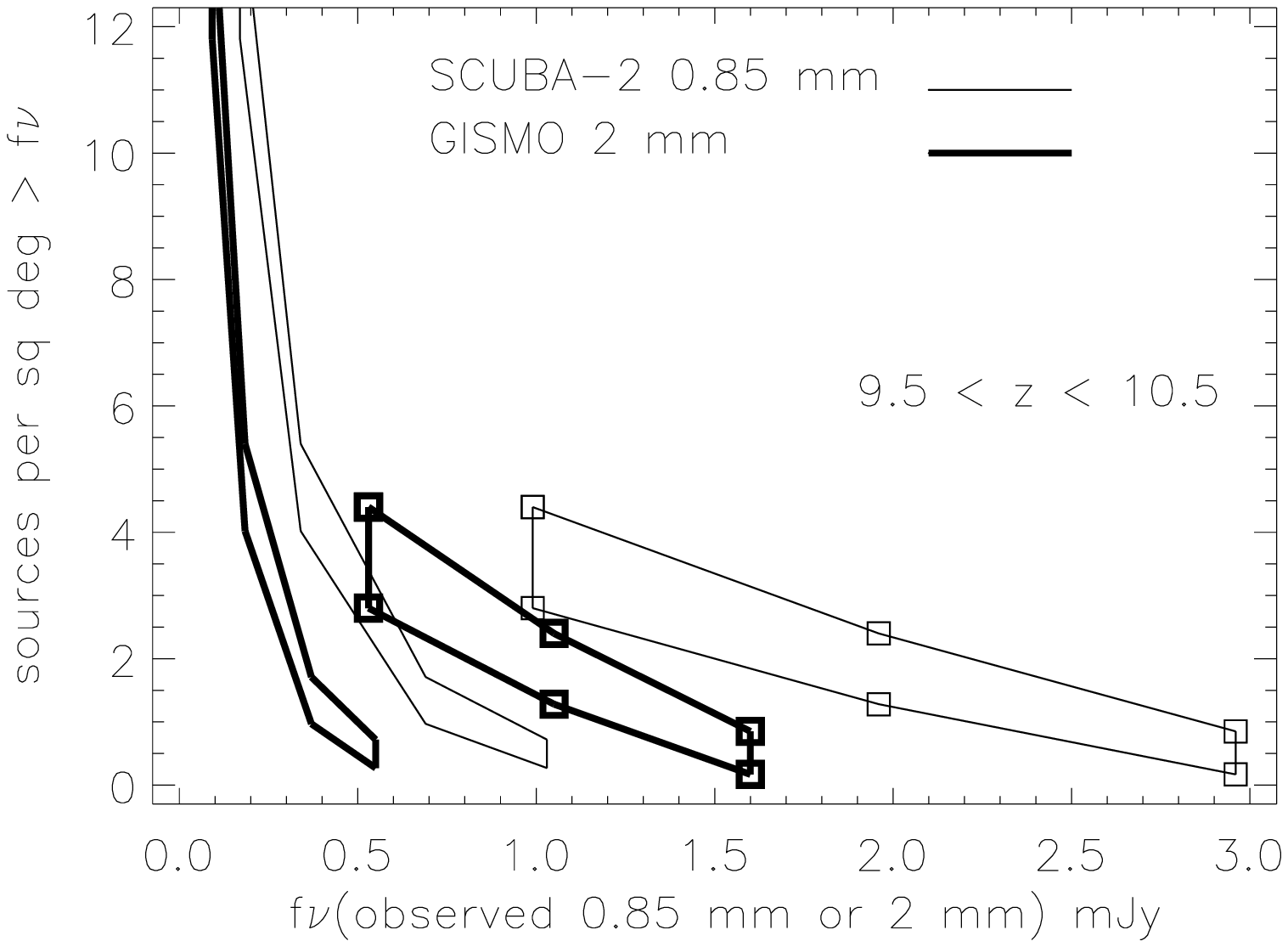}
\caption{Expected quasar counts within 9.5 $<$ z $<$ 10.5 at observed wavelengths 850 \um (SCUBA-2 = thin lines) and 2 mm (GISMO = thick lines) for unobscured and obscured quasars if luminosity functions for 9.5 $<$ z $<$ 10.5 are the same as for 1.8 $<$ z $<$ 2.4 so that 7.8 \um source counts scale from Figure 10 and SEDs scale as in Figures 3 and 4.  Envelopes without symbols are unobscured quasars like those from AGES and envelopes with squares are like obscured IRS quasars in Table 1.  Range of envelopes includes statistical uncertainties in the Bo{\"o}tes counts of these quasar populations. Current SCUBA-2 850 \um detection limit is $\sim$ 2 mJy and GISMO 2mm limit is $\sim$ 0.5 mJy. } 

\end{figure}

\section{Summary and Conclusions}

A population of quasars including all categories of dust obscuration is determined for 1.8 $<$ z $<$ 2.4 using sources in the Bo{\"o}tes and FLS 24 \um survey fields with redshifts from the AGES survey or from the $Spitzer$ IRS.  Luminosities are normalized to rest frame dust luminosities $\nu L_{\nu}$(7.8 \ums), which provides the best comparison among unobscured quasars with 9.7 \um silicate emission and obscured quasars with 9.7 \um absorption.  Obscuration is quantitatively classified by the ratio UV/IR = $\nu L_{\nu}$(0.25 \ums)/$\nu L_{\nu}$(7.8 \ums); unobscured quasars have log UV/IR $>$ 0.2, partially obscured have -1.8 $<$ log UV/IR $<$ 0.2, and obscured have log UV/IR $<$ -1.8.  Quasar counts based on rest frame $f_{\nu}$(7.8 \ums) within the flux density limits of available 24 \um surveys are given for 1.8 $<$ z $<$ 2.4 where it is found that each category of obscuration contributes approximately the same number of quasars (Figure 6). 

SEDs extending to $\sim$ 100 \um are determined using $Herschel$ SPIRE photometry of obscured and unobscured quasars (Figures 3 and 4). New SPIRE photometry is presented for 77 unobscured quasars from the SDSS extending to z = 5.  It is found that the ratio $L_{\nu}$(100 \ums)/$L_{\nu}$(7.8 \ums) is about three times higher for obscured quasars compared to unobscured; the median ratio log [$L_{\nu}$(100 \ums)/$L_{\nu}$(7.8 \ums)] for unobscured quasars is 0.6 but is 1.05 for obscured quasars, a difference of about 2 $\sigma$ compared to the dispersions in the ratios. After correcting for observational uncertainties, the intrinsic variation in the ratio of cool dust to hot dust is $\sim$ $\pm$ 40\% within each category.  Results mean that obscured quasars appear to have a larger fraction of cool dust compared to hot dust, but we cannot determine if this is intrinsic or is caused by differential extinction at 7.8 \um for the obscured quasars.  The $L_{\nu}$(100 \ums)/$L_{\nu}$(7.8 \ums) ratio is less by about a factor of two for quasars compared to local AGN of the same obscuration class. 

Using the far infrared SEDs together with the quasar counts at 7.8 \ums, total quasar counts within 1.8 $<$ z $<$ 2.4 are predicted at observed 450 \um and 850 \um wavelengths for comparison to the submm sources that have been discovered with SCUBA-2 (Figures 7 and 8). It is found that only $\sim$ 5\% of the high redshift submm sources are quasars, and most of these are obscured quasars.  Quasar counts are predicted for 9.5 $<$ z $<$ 10.5 if the luminosity functions and quasar/starburst mix do not change from z = 2 (Figures 10 and 11), and we find that existing SCUBA-2 850 \um surveys or 2 mm surveys with the GISMO survey camera should already have detected sources at z $\sim$ 10 in this case.  This illustrative calculation demonstrates the importance of extending the submm and mm surveys to larger areas and of obtaining redshifts for the unidentified sources in these surveys.

\acknowledgements

We thank those who built the $Herschel$ Observatory for the opportunity to observe with open time. SPIRE was developed by a consortium of institutes led by Cardiff University (UK) and including University of Lethbridge (Canada); NAOC (China); CEA, LAM (France); IFSI, Univ. Padua (Italy); IAC (Spain); Stockholm Observatory (Sweden); Imperial College London, RAL, UCL-MSSL, UKATC, University of Sussex (UK); and Caltech, JPL, NHSC, University of Colorado (USA). This development has been supported by national funding agencies CSA (Canada); NAOC (China); CEA, CNES, CNRS(France); ASI (Italy); MCINN (Spain); SNSB (Sweden);STFC, UKSA (UK); and NASA (USA).  Partial support for this work was provided by NASA through RSA 1489723 issued by JPL/Caltech through the NASA $Herschel$ Science Center.

This publication makes use of data products from the Wide-field Infrared Survey Explorer, which is a joint project of the University of California, Los Angeles, and JPL/Caltech, funded by NASA.

We also acknowledge data products from the SDSS. Funding for the SDSS and SDSS-II has been provided by the Alfred P. Sloan Foundation, the Participating Institutions, the National Science Foundation, the U.S. Department of Energy, the National Aeronautics and Space Administration, the Japanese Monbukagakusho, the Max Planck Society, and the Higher Education Funding Council for England. The SDSS Web site is http://www.sdss.org/. The SDSS is managed by the Astrophysical Research Consortium (ARC) for the Participating Institutions. The participating institutions are the American Museum of Natural History, Astrophysical Institute of Potsdam, University of Basel, Cambridge University, Case Western Reserve University, University of Chicago, Drexel University, Fermilab, the Institute for Advanced Study, the Japan Participation Group, Johns Hopkins University, the Joint Institute for Nuclear Astrophysics, the Kavli Institute for Particle Astrophysics and Cosmology, the Korean Scientist Group, the Chinese Academy of Sciences (LAMOST), Los Alamos National Laboratory, the Max-Planck-Institute for Astronomy (MPIA), the Max-Planck-Institute for Astrophysics (MPA), New Mexico State University, Ohio State University, University of Pittsburgh, University of Portsmouth, Princeton University, the United States Naval Observatory, and the University of Washington.

\clearpage

\begin{deluxetable}{lccccccccc} 
\rotate
\tablecolumns{10}
\tabletypesize{\scriptsize}
\tablewidth{0pc}
\tablecaption{Obscured Quasars Discovered by $Spitzer$ IRS}
\tablehead{
 \colhead{Source} &\colhead{identifier\tablenotemark{a}} &\colhead{coordinates} &\colhead{z\tablenotemark{b}} & \colhead{$f_{\nu}$(7.8 $\mu$m)\tablenotemark{c}}& \colhead{ $\nu L_{\nu}$(7.8 \ums)\tablenotemark{d}}& \colhead{$f_{\nu}$(250 $\mu$m)\tablenotemark{e}} & \colhead{$f_{\nu}$(350 $\mu$m)\tablenotemark{e}} &\colhead{$f_{\nu}$(500 $\mu$m)\tablenotemark{e}} &\colhead{ref.\tablenotemark{f}}
\\
\colhead{} &  \colhead{} &\colhead{J2000}  & \colhead{} & \colhead{mJy} & \colhead{log erg s$^{-1}$} & \colhead{mJy} & \colhead{mJy} & \colhead{mJy} & \colhead{}
}
\startdata

1& SST24(Bo{\"o}tes)    & 142538.23+351855.1	&	2.28 &	1.3& 45.76&	56 &48	&35 &1 \\ 
2& SST24(Bo{\"o}tes) & 142611.35+351217.9 & 1.82	 &	 2.1 & 45.79 & \nodata& \nodata & \nodata &2 \\
3& SST24(Bo{\"o}tes)    & 142622.01+345249.2	&	1.98 &	2.3	&45.90 & $<$20	& $<$20	& $<$25 &3 \\
4& SST24(Bo{\"o}tes)    & 142648.90+332927.2	&	1.82 &	3.3	&45.99 & $<$20	& $<$20	& $<$25 &3 \\
5& SST24(Bo{\"o}tes)    & 142653.23+330220.7	&	1.80 &  1.5 &45.63  &36 &24	& $<$25 &3 \\
6& SST24(Bo{\"o}tes) & 142745.88+342209.0 &3.35	 &	4.5 & 46.58 & \nodata& \nodata & \nodata &2\\
7& SST24(Bo{\"o}tes)    & 142804.12+332135.2	&	2.16 &	1.6 &45.81 &	19	& $<$20	& $<$25  &3\\
8& SST24(Bo{\"o}tes)  & 142924.83+353320.3	&	2.05 &	1.3	&45.67 & $<$20	& $<$20	& $<$25 &1 \\
9& SST24(Bo{\"o}tes) & 142931.36+321828.2 &2.33	 &	1.5 & 45.84 &\nodata& \nodata & \nodata \\
10& SST24(Bo{\"o}tes)    & 142958.33+322615.4	&	2.34 &	1.8	&45.92 & $<$20	& $<$20	& $<$25 &1 \\
11& SST24(Bo{\"o}tes)    & 143001.91+334538.4	 &	2.46 &	5.8 &46.46 &	64 &	55 &	39 &1	\\
12& SST24(Bo{\"o}tes) & 143004.77+340929.9 &3.22	 &	4.1 & 46.51 & \nodata& \nodata & \nodata  &2   \\
13& SST24(Bo{\"o}tes)    & 143025.74+342957.3	 &	2.73 &	3.6 &46.33 &	26 &	21	& $<$25  &3\\
14& SST24(Bo{\"o}tes) & 143026.04+331516.3 & 1.83 & 2.2  & 45.81 & \nodata& \nodata & \nodata &5\\
15& SST24(Bo{\"o}tes)    & 143028.52+343221.3	&	2.15  &	1.9&45.88 &	40 &	37 &	29 &4 \\
16& SST24(Bo{\"o}tes)    & 143109.78+342802.7	&	2.2 &	1.3 &45.73	& $<$20	& $<$20	& $<$25  &3\\
17& SST24(Bo{\"o}tes)    & 143135.29+325456.4 &		1.52 &	4.3 &45.95 &	60 &	55 &	35 &3\\
18& SST24(Bo{\"o}tes)    & 143251.89+333536.8  &		1.70 &1.1  &45.45 &		24 &	18	& $<$25  &3\\
19&  SST24(Bo{\"o}tes) & 143253.39+334844.3 &2.90	 &	1.9 & 46.10 &\nodata& \nodata & \nodata  &2  \\
20& SST24(Bo{\"o}tes) &    143312.70+342011.0	&	2.11 &	2.2 &45.93 &	19	& $<$20	& $<$25  &4 \\
21& SST24(Bo{\"o}tes)&  143318.59+332127.0  & 2.72 	 &	 1.4 	& 45.92 &\nodata& \nodata & \nodata  &2 \\
22& SST24(Bo{\"o}tes)   & 143358.07+332607.7	&	1.95 & 1.7  &45.75 & 20	& $<$20	& $<$25  &3\\
23& SST24(Bo{\"o}tes)    & 143447.70+330230.6	&	1.99 &	2.2 &45.88 &	96 &	69 &	56  &3\\
24& SST24(Bo{\"o}tes)    & 143504.12+354743.2	&	2.08 &	1.6 &45.78 &	22 &	19 & $<$25	 &3\\
25& SST24(Bo{\"o}tes)   & 143508.49+334739.8	&	2.08 &	3.4	&46.10 & $<$20	& $<$20	& $<$25 &3\\
26& SST24(Bo{\"o}tes)    & 143520.75+340418.2	&	2.2 &	2.0 &45.92	& $<$20	& $<$20	& $<$25	 &1\\
27& SST24(Bo{\"o}tes)    & 143523.99+330706.8  &		2.59 &	1.3 &45.85 &	16	& $<$20	& $<$25 &1\\
28& SST24(Bo{\"o}tes)    & 143539.34+334159.1	 &	2.5 & 3.7 &46.28 &	34 &	22	& $<$25  &1\\
29& SST24(Bo{\"o}tes)   & 143545.11+342831.4  &		2.53 &	3.0 &46.20 &	16	& $<$20	& $<$25	 &3\\
30& SST24(Bo{\"o}tes)    & 143644.22+350627.4	&	1.8	 &3.3 &45.98 &	50 &	40 &	27 &1 \\
31& SST24(Bo{\"o}tes)    & 143725.23+341502.4	&	2.04 &	1.9 &45.84 &	33 &	34 &	26  &3\\
32 & SST24(Bo{\"o}tes) & 143807.92+341612.4 & 2.33 & 2.6  & 46.07 & \nodata& \nodata & \nodata  &2 \\
33& SST24(Bo{\"o}tes)    & 143808.34+341015.6	&	2.33 &	2.3	&46.02 & $<$20	& $<$20	& $<$25 &3\\
34& SST24(FLS) & 171057.45+600745.2	&2.34	 &	3.0	 &	46.08 &\nodata& \nodata & \nodata &6\\ 
35 & MIPS8392 & 171343.88+595714.5 & 1.81 & 2.0 & 45.76 & $<$21 & $<$21 & $<$23 &7\\  
36&MIPS532 & 171526.06+585632.7	 &	1.51 &	2.4 &45.66 &	16 &	17	& $<$23 &8\\
37&MIPS8245  &171536.34+593614.8 &		2.65 &	2.0 &46.06	& $<$20	& $<$19	& $<$20 &8\\
38& MIPS78 &171538.18+592540.1	&	2.46 &	3.8 &46.25	& $<$20	& $<$19	& $<$21 &9\\
39 & MIPS8413 & 171545.7+595156.4 & 2.23 & 1.2 & 45.71 & 15 & $<$15 & $<$16 &7 \\
40 & MIPS429  &171611.81+591213.3	&	2.12 &	1.6 &45.80 &	17	& $<$17	& $<$17 &9\\
41 & MIPS42 &171758.44+592816.8	&	2.06 &	5.6 &46.30	& $<$20	& $<$19	& $<$24 &9\\
42&  MIPS22303  &171848.80+585115.1	&	2.34 &	2.7 &46.10	& $<$18	& $<$16	& $<$20 &8\\
43 & SST24(FLS) &  172048.02+594320.6  &2.24	& 	1.9	 &	45.90 & \nodata& \nodata & \nodata &6\\ 
44 &   MIPS16122  &172051.48+600149.1	 &	2.00 &	1.5 &45.72	& $<$19	& $<$16	& $<$18 &8\\
45&MIPS16037 &172133.83+595046.9	 &	1.59 &	2.2 &45.70 &	16	& $<$21	& $<$20 &8 \\
46& MIPS15958 &172324.84+592455.5	&	1.95 &	1.6 &45.74 &	31 &	19	& $<$20 &7\\
47& MIPS22548 &172330.46+584544.9	&	2.21 &	1.5 &45.78	& $<$21	& $<$23	& $<$26 & 8\\
48& SST24(FLS) &  172448.65+601439.9 & 2.34	 &	3.5	 &	46.14  & \nodata& \nodata & \nodata &6\\

\enddata

\tablenotetext{a}{Sources with Bo{\"o}tes numbers correspond to numbers in Table 2 of \citet{var14}.  Sources with MIPS names are from references listed for sources in the $Spitzer$ First Look Survey. }
\tablenotetext{b}{Redshift z measured on CASSIS spectra from fitting median template of AGN with silicate absorption from \citet{sar11}.  Redshifts for Bo{\"o}tes sources are from \citet{var14}; redshifts for MIPS sources are newly measured. Uncertainty in z is $\pm$ 0.08 as determined from scatter in new measures of z compared to original measures using other templates or independent spectra.}
\tablenotetext{c}{Peak flux density at 7.8 \um determined by median of all points in spectrum between 7.7 \um and 7.9 \ums.  Relative uncertainty among sources is $\pm$ 10\% because of poor S/N of faint spectra.  Absolute uncertainty of CASSIS flux calibration applied to all sources is below $\pm$ 3\%. }
\tablenotetext{d}{Rest frame luminosity $\nu L_{\nu}$(7.8 $\mu$m) in erg s$^{-1}$ determined as $\nu L_{\nu}$(7.8 $\mu$m) =  4$\pi$D$_{L}$$^{2}$[$\nu$/(1+z)]$f_{\nu}$(7.8 $\mu$m), for $\nu$ corresponding to 7.8 \ums , taking luminosity distances from \citet{wri06}: www.astro.ucla.edu/~wright/CosmoCalc.html, for H$_0$ = 74 \kmsMpc, $\Omega_{M}$=0.27 and $\Omega_{\lambda}$=0.73. (Log [$\nu L_{\nu}$(7.8 $\mu$m)(\ldot)] = log [$\nu L_{\nu}$(7.8 $\mu$m)(erg s$^{-1}$)] - 33.59.)}
\tablenotetext{e}{SPIRE flux densities and limits from \citet{mel12} for Bo{\"o}tes sources and from \citet{saj12} for MIPS sources. }
\tablenotetext{f}{Reference to original IRS discovery: 1 = \citet{hou05}, 2 = \citet{wee06}, 3 = \citet{bus09}, 4 = \citet{brn07}, 5 = \citet{brn08}, 6 = \citet{wee06b}, 7 = \citet{das09}, 8 = \citet{saj07}, 9 = \citet{yan07}.}

\end{deluxetable}

\clearpage

\begin{deluxetable}{lcccccccccccc} 
\rotate
\tablecolumns{13}
\tabletypesize{\scriptsize}
\tablewidth{0pc}
\tablecaption{Far Infrared Fluxes from SPIRE for High Redshift SDSS Quasars } 
\tablehead{
 \colhead{No.} &\colhead{SDSS identifier\tablenotemark{a}} &\colhead{z\tablenotemark{a}} &\colhead{$f_{\nu}$(22 \ums)\tablenotemark{b}} & \colhead{$f_{\nu}$(7.8 \ums)\tablenotemark{c}}& \colhead{ $\nu L_{\nu}$(7.8 \ums)\tablenotemark{d}}& \colhead{$f_{\nu}$(250 \ums)\tablenotemark{e}} & \colhead{RA,Dec.\tablenotemark{f}} & \colhead{$f_{\nu}$(350 \ums)\tablenotemark{e}} & \colhead{RA,Dec.\tablenotemark{f}} & \colhead{$f_{\nu}$(500 \ums)\tablenotemark{e}} & \colhead{RA,Dec.\tablenotemark{f}}&\colhead{Herschel} 
\\

\colhead{} & \colhead{J2000} &  \colhead{} & \colhead{mJy}& \colhead{mJy} & \colhead{log erg s$^{-1}$} & \colhead{mJy} & \colhead{arcsec} & \colhead{mJy} & \colhead{arcsec}& \colhead{mJy} & \colhead{arcsec}&\colhead{i.d.}
}
\startdata

1 & 030341.04-002321.9 & 3.18 &6.2 & 8.3 & 46.82 & 31.9 & -0.2,-1.8 & $<$25 & 0.3,56.3 & $<$25 & 56.0,-48.9 & 1342224969	\\		 
2 & 032108.45+413220.8 & 2.47 &12.7& 14.8 & 46.87 & $<$25 & 686,190 & $<$25 & 609,49.6 & $<$25 & 763,-19.7 & 1342203614	\\		 
3 & 073100.16+430740.6 & 3.95 &4.5& 6.8 & 46.87 & $<$25 & 1.1,-44.5 & $<$25 & 31.6,-18.4 & $<$25 & -72.2,-44.2 & 1342270267	\\		 
4 & 073103.12+445949.4 & 5.00 &2.2& 4.1 & 46.80 & 22.2 & 4.8,15.8 & 15.5 & 2.3,14.9 & $<$25 & -38.1,-46.8 & 1342204959	\\		 
5 & 073502.30+265911.5 & 1.97 &27.6& 28.8 & 46.99 & 84.8 & -2.3,2.0 & 45.0 & 1.3,0.3 & 31.7 & 4.2,-2.7 & 1342270324	\\		 
6 & 073733.01+392037.4 & 1.74 &33.7& 32.9 & 46.94 & 119.8 & -1.7,1.4 & 71.7 & -1.1,2.3 & 43.6 & -0.1,2.2 & 1342270269	\\		 
7 & 073936.25+280754.7 & 2.74 &6.9& 8.5 & 46.71 & $<$25 & -70.7,-55.2 & 23.3 & -6,2.7 & 37.8 & -14.5,8.2 & 1342270323	\\		 
8 & 074126.07+421530.7 & 1.85 &16.0& 16.2 & 46.68 & 67.9 & -0.9,-1.3 & 31.5 & 4.7,-2 & $<$25 & 74.6,53.8 & 1342270268	\\		 
9 & 074521.78+473436.1 & 3.22 &11.1& 14.9 & 47.06 & 40.9 & 1.8,-4.7 & 46.1 & 3.9,-3.9 & 50.0 & 2.1,-2.7 & 1342268353	\\		 
10 & 074815.82+355912.2 & 3.36 &6.4& 8.9 & 46.88 & 77.9 & -0.1,1.2 & 83.7 & 0.3,1.9 & 50.2 & 1.4,5.6 & 1342270272	\\		 
11 & 074914.78+472904.1 & 2.04 &13.8& 14.6 & 46.72 & 74.4 & -1.9,0.3 & 60.5 & -0.1,3.1 & $<$25 & -57.2,8.9 & 1342270266	\\		 
12 & 075054.64+425219.2 & 1.90 &22.2& 22.7 & 46.85 & $<$25 & 38.7,11.9 & $<$25 & -77.1,-28.3 & $<$25 & -102,251 & 1342270273	\\		 
13 & 075732.89+441424.6 & 4.17 &3.1& 4.9 & 46.76 & $<$25 & -29.2,-10.0 & $<$25 & 62.8,10.6 & $<$25 & 105,-28.1 & 1342270274	\\		 
14 & 080117.79+521034.5 & 3.24 &12.3& 16.6 & 47.12 & 79.8 & -2.9,0.2 & 79.7 & -0.3,0.4 & 56.7 & -2.2,-5 & 1342270265	\\		 
15 & 080542.39+155528.1 & 2.51 &7.70& 9.5 & 46.69 & 35.6 & -1.1,3.6 & $<$25 & -7.8,29.5 & 62.6 & -4.7,30.6 & 1342270317	\\		 
16 & 080849.42+521515.3 & 4.46 &3.1& 5.1 & 46.82 & 30.8 & -1.5,1.1 & 40.9 & -6,-3.5 & 46.2 & -8.3,1.9 & 1342270264	\\		 
17 & 081114.66+172057.4 & 2.30 &10.8& 12.2 & 46.73 & $<$25 & -40.9,-21.1 & $<$25 & -47.4,-59.3 & $<$25 & 16.7,49.7 & 1342270318	\\		 
18 & 081207.57+052341.1 & 1.88 &16.5& 16.8 & 46.71 & 81.8 & 2.7,8.8 & 76.0 & 3.5,7.2 & 67.2 & 7.8,5.9 & 1342270309	\\		 
19 & 081331.28+254503.0 & 1.51 &89.5& 80.1 & 47.21 & 188.8 & 1.0,-1.7 & 84.6 & -1.8,-0.4 & $<$25 & -98.6,35.4 & 1342254473	\\		 
20 & 081806.87+071920.2 & 4.58 &3.1& 5.2 & 46.85 & $<$25 & -69.1,4.4 & $<$25 & 63.2,58.6 & $<$25 & 74.3,-57.7 & 1342270310	\\		 
21 & 081855.77+095848.0 & 3.67 &6.5& 9.5 & 46.96 & 47.9 & -3.5,2.4 & 45.2 & -3.5,4.2 & 29.3 & -4.5,4.6 & 1342270312	\\		 
22 & 081940.58+082357.9 & 3.21 &6.3& 8.4 & 46.82 & 37.3 & 1.7,3.4 & $<$25 & -50.7,8.7 & $<$25 & -47.6,0.2 & 1342270311	\\		 
23 & 082319.65+433433.7 & 1.66 &35.2& 33.4 & 46.91 & 61.3 & -2.0,0.1 & 40.8 & -2.7,1.4 & $<$25 & 25.0,-71.6 & 1342270277	\\		 
24 & 082450.79+154318.4 & 1.87 &19.5& 19.9 & 46.78 & 132.4 & -1.0,1.2 & 100.5 & -1.8,2.1 & 57.9 & -2.4,2.5 & 1342270315	\\		 
25 & 082454.02+130217.0 & 5.19 &2.2& 4.1 & 46.83 & 28.5 & 0.9,1.8 & 30.0 & -1.2,3.3 & 43.2 & -0.9,-0.8 & 1342270314	\\		 
26 & 082548.07+095339.4 & 3.80 &4.6& 6.9 & 46.84 & $<$25 & 66.6,-92.2 & $<$25 & -61.4,-47.4 & $<$25 & 19.2,-91.9 & 1342270313	\\		 
27 & 082619.70+314847.9 & 3.09 &6.6& 8.7 & 46.80 & 34.3 & -6.9,13.0 & 41.3 & -4.9,11 & 36.1 & 0.2,0.1 & 1342270291	\\		 
28 & 082638.59+515233.2 & 2.85 &11.2& 14.1 & 46.95 & 71.7 & -1.5,0.0 & 55.4 & -3.3,-0.8 & 32.2 & 1.8,-1.6 & 1342270262	\\		 
29 & 082804.54+445256.9 & 2.07 &12.1& 13.4 & 46.69 & 37.7 & -0.3,2.6 & 33.0 & -2.1,4.7 & $<$25 & 3.5,152.5 & 1342270276	\\		 
30 & 082854.70+431220.1 & 3.17 &6.5& 8.6 & 46.82 & 44.4 & -2.2,1.7 & $<$25 & 53.6,-14.5 & $<$25 & -79.3,17.9 & 1342270278	\\		 
31 & 083103.01+523533.5 & 4.44 &2.6& 4.3 & 46.75 & $<$25 & -29.2,-62.2 & $<$25 & -56.5,33.2 & $<$25 & 30.2,-129.7 & 1342270261	\\		 
32 & 083212.37+530327.3 & 4.05 &4.0& 6.2 & 46.85 & 36.7 & -0.3,2.7 & 23.2 & 1.9,0.7 & 30.3 & -0.7,-8.3 & 1342270260	\\		 
33 & 083249.39+155408.6 & 2.42 &8.25& 10.0 & 46.68 & 34.7 & -4.0,0.2 & $<$25 & -16.6,-46.6 & $<$25 & -19.2,-53.1 & 1342270302	\\		 
34 & 083255.63+182300.6 & 2.27 &10.3& 11.6 & 46.70 & $<$25 & -9.4,36.4 & $<$25 & -8.5,33.8 & 27.9 & -1.8,24.2 & 1342270300	\\	 
35 & 083413.90+511214.6 & 2.39 &53.8& 62.0 & 47.47 & 182.0 & -0.6,1.1 & 124.1 & -0.9,1.5 & 56.2 & -3.0,4.5 & 1342270259	\\		 
36 & 083417.12+354833.1 & 2.16 &12.9& 14.2 & 46.74 & 68.3 & 0.3,-1.5 & 77.3 & -0.7,0.7 & 43.9 & -0.2,0.7 & 1342230752	\\	 
37 & 083535.69+212240.1 & 3.12 &8.1& 10.6 & 46.90 & $<$25 & -39.6,-15.8 & $<$25 & -43.3,-16.3 & $<$25 & -52.0,-5.6 & 1342270296	\\		 
38 & 083552.62+163343.9 & 4.25 &9.2& 14.7 & 47.26 & 41.9 & -0.4,-0.4 & 37.2 & 4.9,-1.5 & 48.6 & 3.3,0.7 & 1342270301	\\		 
39 & 083839.16+285852.7 & 4.36 &4.3& 7.0 & 46.95 & $<$25 & 77.3,-39.5 & $<$25 & -97.9,-102.5 & $<$25 & -22.9,138 & 1342270292	\\		 
40 & 083850.15+261105.4 & 1.61 &20.5& 19.5 & 46.66 & $<$25 & -35.1,-8.2 & $<$25 & 34.1,69.3 & $<$25 & 30.5,66.7 & 1342270293	\\		 
41 & 084045.40+090809.4 & 4.54 &2.7& 4.6 & 46.79 & $<$25 & 50.4,-47.9 & 26.4 & -4.5,1.7 & $<$25 & -17.0,56.3 & 1342270305	\\		 
42 & 084051.22+404806.7 & 4.42 &2.5& 4.1 & 46.73 & $<$25 & -50.7,18.9 & 23.0 & 8,3.7 & $<$25 & 42.8,-115 & 1342270279	\\		 
43 & 084401.95+050357.9 & 3.35 &8.6& 11.9 & 46.99 & $<$25 & 63.1,17.7 & $<$25 & 52.1,15.8 & $<$25 & -137,-129 & 1342270306	\\		 
44 & 084438.04+584825.5 & 4.77 &2.5& 4.3 & 46.80 & $<$25 & 61.0,-42.7 & $<$25 & -9.6,24.3 & $<$25 & 126,-123 & 1342270244	\\		 
45 & 084547.19+132858.1 & 1.88 &16.4& 16.7 & 46.71 & 27.5 & -8.8,2.3 & 39.3 & -15.7,1.7 & 43.6 & -15.4,-1.9 & 1342270303	\\		 
46 & 085010.26+593118.2 & 1.72 &25.4& 24.7 & 46.81 & 36.3 & -6.6,4.0 & 23.6 & -12.1,-15.3 & $<$25 & 215,-31 & 1342270243	\\		 
47 & 085210.88+535948.9 & 4.22 &4.6& 7.3 & 46.94 & 39.0 & 12.6,-12.4 & 44.8 & 9.9,-9.8 & 39.8 & 14.5,-15.1 & 1342270246	\\		 
48 & 085335.74+185446.5 & 2.15 &13.3& 14.5 & 46.76 & 31.4 & -2.3,7.0 & 47.3 & -6,21.5 & 47.4 & -4.2,8.7 & 1342270299	\\		 
49 & 085611.69+411516.8 & 3.68 &5.5& 8.0 & 46.89 & $<$25 & 4.4,-22.1 & $<$25 & -39.4,21.8 & $<$25 & -38.3,25.2 & 1342270280	\\		 
50 & 085626.47+194137.7 & 2.82 &30.4& 38.1 & 47.38 & $<$25 & -64.5,-6.1 & $<$25 & -37.7,58.4 & $<$25 & 101,-114 & 1342270298	\\		 
51 & 085634.92+525206.2 & 4.82 &1.9& 3.3 & 46.69 & 36.2 & -13.2,5.4 & 49.3 & -9.6,6 & 32.6 & -4.1,1.9 & 1342270248	\\		 
52 & 085707.94+321031.9 & 4.78 &2.6& 4.5 & 46.82 & $<$25 & 17.9,-21.6 & $<$25 & 17.2,19.6 & $<$25 & -48.4,-106 & 1342230758	\\		 
53 & 090033.50+421547.0 & 3.29 &9.6& 13.1 & 47.02 & 30.2 & -3.0,2.5 & 35.3 & 0,-2 & $<$25 & -40.0,-53.8 & 1342270281	\\		 
54 & 090158.85+610931.7 & 4.08 &3.5& 5.4 & 46.79 & 29.1 & -8.5,-5.8 & 31.7 & -13.4,-4.2 & 32.8 & -22.0,-4.5 & 1342270241	\\		 
55 & 090334.94+502819.3 & 3.58 &8.1& 11.6 & 47.03 & 220.9 & -0.3,-1.3 & 237.8 & 0.5,-1.7 & 189.9 & 0.7,-2.3 & 1342254628	\\		 
56 & 090527.46+485049.9 & 2.69 &8.2& 10.0 & 46.76 & $<$25 & 118.8,8.3 & 27.7 & 13.3,16.4 & $<$25 & -112,31.6 & 1342270257	\\		 
57 & 091206.78+331109.3 & 3.33 &7.0& 9.6 & 46.90 & 57.0 & -1.9,0.2 & 45.7 & -2.6,-1.1 & 28.9 & -14.1,1 & 1342270288	\\		 
58 & 091301.01+422344.7 & 2.31 &10.8& 12.3 & 46.74 & 45.4 & -0.8,5.4 & 24.5 & 3.7,-1.5 & $<$25 & -31.5,-30.9 & 1342270282	\\		 
59 & 091342.48+372603.3 & 2.13 &12.1& 13.2 & 46.70 & 36.1 & -2.1,-0.6 & 33.1 & 1.6,4 & 37.3 & -6.8,6.1 & 1342270284	\\		  
60 & 091610.35+621326.2 & 2.08 &21.3& 22.9 & 46.92 & 119.6 & -1.8,2.4 & 75.9 & -1.6,3.3 & 55.1 & 1.2,2.5 & 1342270239	\\		 
61 & 092058.46+444154.0 & 2.19 &19.3& 21.3 & 46.93 & 112.7 & 0.3,1.2 & 164.5 & 0,0 & 193.9 & 0.9,-0.8 & 1342270255	\\		 
62 & 092819.29+534024.1 & 4.39 &5.0& 8.2 & 47.02 & 56.5 & 4.1,2.5 & 66.3 & 2.7,4.6 & 56.8 & -0.1,3.5 & 1342270249	\\		 
63 & 093554.46+525616.4 & 4.01 &5.0& 7.6 & 46.93 & $<$25 & -36.6,-14.7 & $<$25 & -36.3,-11 & $<$25 & -38.0,176 & 1342270250	\\		 
64 & 094056.01+584830.2 & 4.66 &3.1& 5.4 & 46.88 & 29.2 & 1.5,8.0 & 29.2 & -0.8,5.2 & $<$25 & -47.2,-27.4 & 1342270236	\\		 
65 & 095014.05+580136.5 & 3.96 &4.4& 6.7 & 46.86 & 39.5 & -14.1,-4.2 & 45.0 & -13.1,-8.4 & 30.4 & -10.4,-8.1 & 1342270235	\\		 
66 & 100129.64+545438.1 & 1.76 &15.8& 16.0 & 46.64 & 38.8 & -11.1,7.6 & 48.2 & -11.9,5.1 & 40.1 & -11.1,7.8 & 1342270251	\\		 
67 & 101051.14+570530.8 & 1.96 &11.9& 12.8 & 46.63 & 40.6 & -1.6,5.5 & 30.2 & -7.9,13.5 & $<$25 & -89.5,35.1 & 1342270233	\\		 
68 & 102907.09+651024.6 & 2.16 &12.3& 13.4 & 46.73 & 26.5 & -0.6,1.3 & 28.4 & 2.4,-0.1 & $<$25 & -96.3,-64.3 & 1342270222	\\		 
69 & 140146.53+024434.7 & 4.44 &2.9& 4.7 & 46.79 & $<$25 & 318,-58.6 & $<$25 & 288,-251 & $<$25 & 194,-326 & 1342201130	\\		 
70 & 141546.24+112943.4 & 2.56 &57.1& 68.0 & 47.56 & 531.8 & 1.1,-0.4 & 399.4 & 1,-0.5 & 229.3 & 0.5,0 & 1342261537	\\		 
71 & 144709.24+103824.5 & 3.68 &8.3& 12.1 & 47.07 & $<$25 & -309,394 & $<$25 & -225,447 & $<$25 & -261,513 & 1342236153	\\		 
72 & 150424.98+102939.1 & 1.84 &18.6& 18.7 & 46.74 & 148 & 0.8,-2.5 & $<$25 & 53.3,-3.8 & $<$25 & 350,254 & 1342238323	\\		 
73 & 153308.65+301820.7 & 4.45 &2.7& 4.5 & 46.77 & 37.3 & 2.2,-0.9 & 29.8 & 0.9,-3.6 & 28.7 & 4.9,-4.7 & 1342261681	\\		 
74 & 160336.64+350824.3 & 4.46 &2.4& 4.0 & 46.73 & 43.7 & 1.2,0.1 & 57.2 & 1.9,-0.2 & 34.8 & 6.6,1.5 & 1342241162	\\		 
75 & 161622.10+050127.7 & 4.87 &2.2& 3.8 & 46.76 & 45.5 & 5.6,-14.8 & 62.2 & 2.8,-13.2 & 41.5 & -3.9,-10.7 & 1342229564	\\		 
76 & 163411.82+215325.0 & 4.53 &3.0& 5.0 & 46.83 & $<$25 & 4.6,32.2 & 21.8 & 1.7,-3.2 & 14.8 & 1.3,-1.8 & 1342239981	\\		 
77 & 172413.27+571046.7 & 2.83 &7.5& 9.4 & 46.77 & 34.5 & -0.2,-8.8 & 44.6 & 2.7,-11.9 & 29.1 & -2.0,-18.1 & 1342270212	\\		 
\enddata
\tablenotetext{a}{SDSS identifier and redshift from version 7 of the SDSS quasar catalog \citep{sch10}.}
\tablenotetext{b}{Observed flux density at 22 \um from the WISE All Sky Catalog available at wise2.ipac.caltech.edu/docs/release/allsky/.  Zero point of 22 \um magnitude listed in catalog taken as 8280 mJy; typical uncertainties for sources with fluxes listed are $\pm$ 15\%.  }
\tablenotetext{c}{Flux density $f_{\nu}$(7.8 \ums) at observed wavelength corresponding to rest wavelength 7.8 \ums, determined by scaling $f_{\nu}$ (observed 22 \ums) to $f_{\nu}$(rest frame 7.8 \ums) using tabulated redshift and template spectrum of silicate emission quasars in \citet{wee12}.} 
\tablenotetext{d}{Rest frame luminosity $\nu L_{\nu}$(7.8 \ums) in erg s$^{-1}$ determined as $\nu L_{\nu}$(7.8 \ums) =  4$\pi$D$_{L}$$^{2}$[$\nu$/(1+z)]$f_{\nu}$(7.8 \ums), for $\nu$ corresponding to 7.8 \ums, taking luminosity distances from \citet{wri06}:  www.astro.ucla.edu/~wright/CosmoCalc.html, for H$_0$ = 74 \kmsMpc, $\Omega_{M}$=0.27 and $\Omega_{\lambda}$=0.73. }
\tablenotetext{e}{Flux density from SPIRE in observed frame at wavelength listed. Median one sigma uncertainty for $f_{\nu}$(250 $\mu$m) is $\pm$ 8.5 mJy, for $f_{\nu}$(350 $\mu$m) is $\pm$ 7.3 mJy, and for $f_{\nu}$(500 $\mu$m) is $\pm$ 9.0 mJy.}
\tablenotetext{f}{Offset in arcsec from SDSS coordinate of closest SPIRE source at wavelength of preceding column.  SPIRE source is identified with the SDSS source and a value for $f_{\nu}$ listed in the preceding column if the total offset distance $<$ 19\arcsec~ at 250 \um, $<$ 25\arcsec~ at 350 \um, and $<$ 37\arcsec~ at 500 \um (i.e., within one FWHM of the beam size.).  If no SPIRE source is found within these distances, flux of the SDSS quasar is listed as upper limit of $<$ 25 mJy in preceding column.}  

\end{deluxetable}

\end{document}